\documentstyle[12pt,epsfig,rotate]{article}

\newcommand{\RE}{{\rm Re}}
\newcommand{\IM}{{\rm Im}}

\newcommand{\qslash}{/ \mskip-8mu q} 
\def\diag{\mathop{\mbox{diag}}}

\newcommand{\nn}{\nonumber}

\newcommand{\mc}{m_{\rm c}}

\newcommand{\bea}{\begin{eqnarray}}
\newcommand{\eea}{\end{eqnarray}}
\newcommand{\bd}{\begin{displaymath}}
\newcommand{\ed}{\end{displaymath}}

\newcommand{\beq}{\begin{equation}}
\newcommand{\eeq}{\end{equation}}
\newcommand{\be}{\begin{equation}}
\newcommand{\ee}{\end{equation}}
\newcommand{\ord}{{\cal O}}

\def\kpnn{$K^+\rightarrow\pi^+\nu\bar\nu$}
\def\kpn{K^+\rightarrow\pi^+\nu\bar\nu}
\def\klpn{K_{\rm L}\rightarrow\pi^0\nu\bar\nu}
\def\klpnn{$K_{\rm L}\rightarrow\pi^0\nu\bar\nu$}

\begin{document}
\begin{flushright}
 TUM-HEP-302/97 \\
 SNS/PH/1997-9 \\
 hep-ph/9712398 \\
 December 1997
\end{flushright}
\vskip1truecm
\centerline{\LARGE\bf  $K\to\pi\nu\bar\nu$: A Model Independent}
\centerline{\LARGE\bf  Analysis and Supersymmetry}
   \vskip1truecm
\centerline{\large\bf Andrzej J. Buras$^a$, Andrea Romanino$^{a,b}$ and
Luca Silvestrini$^a$}
\bigskip
\centerline{\sl $^a$Technische Universit\"at M\"unchen, Physik Department}
\centerline{\sl D-85748 Garching, Germany}
\centerline{\sl $^b$Scuola Normale Superiore and INFN, sezione di Pisa} 
\centerline{\sl I-56126 Pisa, Italy}
\vskip1truecm
\thispagestyle{empty}
\centerline{\bf Abstract}

We present a model independent analysis of new-physics contributions
to the rare decays $\kpn$ and $\klpn$. We parameterize the effects of 
new physics in
these decays by two parameters: $r_K$ and the phase $\theta_K$, with
$r_K=1$ and $\theta_K=0$ in the Standard Model. We show how these
parameters can be extracted from future data together with the
relevant CKM parameters, in particular the angle $\beta$ of the
unitarity triangle.  To this end CP asymmetries in $B \to \psi K_S$
and $B \to \pi^+\pi^-$ as well as the ratio $|V_{ub}/V_{cb}|$ have to
be also considered. This analysis offers simultaneously some insight
in a possible violation of a ``golden relation'' between $K \to
\pi\nu\bar\nu$ decays and the CP asymmetry in $B \to \psi K_S$ in the
Standard Model pointed out some time ago. We illustrate these ideas by
considering a general class of supersymmetric models. We find that in
the ``constrained'' MSSM, in which $\theta_K=0$, the measurements of
Br$(\kpn)$ and Br$(\klpn)$ directly determine the angle $\beta$.
Moreover, the ``golden relation'' remains unaffected.  On the other
hand, in general SUSY models with unbroken R-parity the present
experimental constraints still allow for substantial deviations from
$r_K=1$ and $\theta_K=0$.  Typically $0.5 < r_K < 1.3$ and $-25^0 <
\theta_K < 25^0$. Consequently, in these models the violation of the
``golden relation'' is possible and values for Br$(\kpn)$ and
Br$(\klpn)$ departing from the Standard Model expectations by factors
2--3 cannot be excluded. Simultaneously, the extraction of the
``true'' angle $\beta$ from $K \to \pi \nu \bar \nu$ is not possible
without additional information from other decays. Our conclusions
differ in certain aspects from the ones reached in previous analyses.
In particular, we stress the possible importance of left-right
flavour-violating mass insertions that were not considered before.

\setcounter{page}{1}
\newpage
\section{Introduction}
\label{sec:intro}
Among the CP asymmetries in $B$ decays, the CP asymmetry in
$B_d\to\psi K_S$ \cite{BS81} is unique as it is the only one which in
the Standard Model measures directly one angle of the
unitarity triangle, the angle $\beta$, without any hadronic
uncertainties.

On the other hand, it has been pointed out in \cite{BB96} that a
similar role among the $K$ decays is played by  
$K_L\to\pi^0\nu\bar\nu$ and $K^+\to\pi^+\nu\bar\nu$.
Indeed, taken together, these two decays offer a very clean determination
of $\sin 2\beta$ by measuring their branching ratios only \cite{BB96}:

\begin{equation}\label{sin}
\sin 2\beta=\frac{2 r_s}{1+r^2_s}\,,
\end{equation}
where
\begin{equation}\label{cbb}
r_s\equiv
r_s(B_1, B_2)=\cot\beta=
\sqrt{\sigma}{\sqrt{\sigma(B_1-B_2)}-P_c\over\sqrt{B_2}}\,.
\end{equation}
Here $\sigma=1/(1-\lambda^2/2)^2$ with $\lambda=0.22$,
\begin{equation}\label{b1b2}
B_1={{\rm Br}(\kpn)\over 4.11\cdot 10^{-11}}\,,\qquad
B_2={{\rm Br}(\klpn)\over 1.80\cdot 10^{-10}}\,,
\end{equation}
and $P_c=0.40\pm 0.06$ represents the internal charm contribution to
the amplitude $A(K^+\to\pi^+\nu\bar\nu)$ which is known including
next-to-leading QCD corrections \cite{BB13}. The error in $P_c$ is
dominated by the renormalization scale uncertainties and the value
of $\mc$.

Using the well-known expression for $\sin 2 \beta$ in terms of the
time-integrated CP asymmetry
$a_{ \psi K_S}$ in  $B_d\to\psi K_S$ decay,
 one finds an interesting connection
between rare $K$ decays and $B$ physics \cite{BB96}:
\begin{equation}\label{kbcon}
{2 r_s(B_1,B_2)\over 1+r^2_s(B_1,B_2)}=
-a_{\psi K_ S}
{1+x^2_d\over x_d}\,,
\end{equation}
or equivalently
\be\label{gr}
[\sin 2\beta]_{\pi\nu\bar\nu}=[\sin 2\beta]_{\psi K_{\rm S}}\,,
\ee
which must be satisfied in the Standard Model. 
Here $x_d$ measures the $B_d^0-\bar B_d^0$ mixing.
As stressed in \cite{BB96}, this ``golden relation'' involves,
except
for $P_c$, only  directly measurable quantities. 
Due to very small theoretical uncertainties in (\ref{kbcon}), this
relation is particularly suited for tests of CP violation in the
Standard Model and offers a powerful tool to probe the physics
beyond it.
These points have been recently reemphasized in \cite{GN,GN1}.

The present status of the Standard Model predictions for $K \to \pi \nu
\bar \nu$ can be
summarized by \cite{BF97}:
\bea
{\rm Br} (K^+ \to \pi^+ \nu \bar \nu) = (9.1 \pm 3.8 ) \cdot 10^{-11},
\nonumber \\
{\rm Br} (K_L \to \pi^0 \nu \bar \nu) = (2.8 \pm 1.7 ) \cdot 10^{-11},
\label{brsm}
\eea
where the errors are dominated by the uncertainties in the CKM
parameters. On the experimental side, the first event for $\kpn$   
has been recently observed by the BNL787 collaboration \cite{bnl787},
giving
\be
{\rm Br} (K^+ \to \pi^+ \nu \bar \nu) = (4.2^{+9.7}_{-3.5} ) \cdot
10^{-10}, 
\ee
in the ball park of the Standard Model expectations. The most recent
upper bound on Br(\klpnn) from FNAL-E799 is $1.8 \cdot 10^{-6}$. 
A new proposal
at Brookhaven, AGS2000 \cite{ags}, expects to reach the single event
sensitivity $2 \cdot 10^{-12}$, allowing a 10\% measurement of
Br(\klpnn).
With a similar accuracy for Br(\kpnn) a measurement of $\sin 2 \beta$
through (\ref{sin}) with an error of $\pm 0.05$ is possible \cite{BB96}. 
This is comparable with the expected accuracy for $\sin 2 \beta$ from
$a_{\psi K_S}$ in the first years of the next decade at $B$
factories. 

The purpose of this paper is to investigate how the relation 
(\ref{kbcon}) could be affected by new physics beyond the
Standard Model. In particular we will consider a large class of
supersymmetric models. This will also allow us to reach
some general conclusions on supersymmetry effects in
$K_L\to\pi^0\nu\bar\nu$ and $K^+\to\pi^+\nu\bar\nu$.

Our paper is organized as follows. In section 2 we briefly recall the
derivation of the formulae above in the manner useful for
generalizations. In section 3 a model independent analysis of 
new-physics effects in $K\to \pi\nu\bar\nu$ decays is presented.  In order
to obtain some insight in the violation of the relation (\ref{kbcon}),
this analysis is then combined with the model independent analysis of
the unitarity triangle presented in \cite{GNW}.  In section 4,
$K_L\to\pi^0\nu\bar\nu$ and $K^+\to\pi^+\nu\bar\nu$ are analyzed in a
large class of supersymmetric models. This analysis gives some insight
in the violation of the relation (\ref{kbcon}) in these models.
There we compare our results with a recent analysis of Nir and Worah
\cite{nir}. Our conclusions differ in certain aspects from the ones
reached by these authors.
Section 5 summarizes the main results of our paper.

\section{The Case of the Standard Model}
\label{sec:SM}

It is instructive to recall first how (\ref{sin}) is derived.
To this end let us  consider the effective Hamiltonian 
for $K \to\pi\nu\bar\nu$:
\begin{equation}\label{hkpn} 
{\cal H}^{\rm SM}_{\rm eff}={G_{\rm F} \over{\sqrt 2}}{2 \alpha\over \pi 
\sin^2\Theta_{\rm W}}
 \sum_{l=e,\mu,\tau}\left(\lambda_c X^l_{\rm NL}+
\lambda_t X(x_t)\right)
 \bar{s}_L \gamma^\mu d_L \bar{\nu}^l_L \gamma_\mu \nu^l_L + {\rm H.c.} \, ,
\end{equation}
which originates in Z-penguin and box diagrams with internal charm 
($X_{NL}$) and top quark ($X(x_t)$) exchanges.
Here $\lambda_i=V^{\ast}_{is}V_{id}$ and $x_t=m_t^2/M_W^2$.
The dependence on the charged lepton mass resulting from the relevant
box-graph
is negligible for the top contribution.
In the charm sector this is the
case only for the electron and the muon but not for the $\tau$-lepton.

Now $K_L\to\pi^0\nu\bar\nu$ is purely
CP violating in the Standard Model with CP violation proceeding 
dominantly in the decay
amplitude \cite{Littenberg}. 
Let us then introduce the quantity
\be\label{FGE}
F=\frac{1}{\lambda^5} \left[ \lambda_c X_{\rm NL}+
\lambda_t X(x_t)\right]\,,
\ee
where
\be
X_{\rm NL}=\frac{2}{3} X^e_{\rm NL} +\frac{1}{3} X^\tau_{\rm NL}
\equiv \lambda^4 P_c
\ee
and
\be\label{XT}
X(x_t)=\eta_X {{x_t}\over{8}}\;\left[{{x_t+2}\over{x_t-1}} 
+ {{3 x_t-6}\over{(x_t -1)^2}}\; \ln x_t\right]\, , 
\ee
with $\eta_X=0.985$ being the QCD correction calculated in \cite{BB13}.
Expressing the matrix elements $<\pi^0|\bar{s}_L \gamma^\mu d_{L}|K_L>$ 
and $<\pi^+|\bar{s}_L \gamma^\mu d_{L}|K^+>$ through 
${\rm Br}(K^+\to\pi^0 e^+\nu)$ one finds:
\begin{equation}\label{bkpn}
{\rm Br}(\kpn)=\kappa_+\cdot\left[ (\RE F)^2+ (\IM F)^2 \right]
\end{equation}
and
\begin{equation}\label{bklpn}
{\rm Br}(K_{\rm L}\to\pi^0\nu\bar\nu)=\kappa_{\rm L}\cdot (\IM F)^2.
\end{equation}
Here
\begin{equation}\label{kapp}
\kappa_+=r_{K^+}{3\alpha^2 {\rm Br}(K^+\to\pi^0e^+\nu)\over 2\pi^2
\sin^4\Theta_{\rm W}}
 \lambda^8=4.11\cdot 10^{-11}
\end{equation}
and
\begin{equation}\label{kapl}
\kappa_{\rm L}=\frac{r_{K_{\rm L}}}{r_{K^+}}
 {\tau(K_{\rm L})\over\tau(K^+)}\kappa_+ =1.80\cdot 10^{-10}\,,
\end{equation}
where we have used
\begin{equation}\label{alsinbr}
\alpha=\frac{1}{129},\qquad \sin^2\Theta_{\rm W}=0.23, \qquad
{\rm Br}(K^+\to\pi^0e^+\nu)=4.82\cdot 10^{-2}\,.
\end{equation}
Finally, $r_{K^+}=0.901$ and $r_{K_{\rm L}}=0.944$ summarize isospin
breaking corrections in relating $\kpn$ 
and $\klpn$ to $K^+\to\pi^0e^+\nu$ respectively
\cite{MP}. The known potential ambiguity in the value for
$\sin^2\Theta_{\rm W}$ can be reduced by including two-loop
electroweak contributions to $K \to \pi \nu \bar \nu$
\cite{BB2l}. With $\sin^2\Theta_{\rm W}=0.23$, these two-loop
electroweak corrections are estimated to be negligible: at most $1-2$\%.

Given ${\rm Br}(\kpn)$ and ${\rm Br}(\klpn)$ one finds:
\be\label{MAIN}
\RE F = -\varepsilon_1 \sqrt{B_1-B_2}, \qquad 
\IM F= \varepsilon_2\sqrt{B_2},
\ee
with the reduced branching ratios defined in (\ref{b1b2}).
$\varepsilon_i$ being equal to $\pm$ indicate the four-fold
discrete ambiguity which must be resolved by using other
decays. 

\begin{figure}   
    \centerline{
    \epsfxsize=6truecm
    \rotate[r]{
    \epsffile{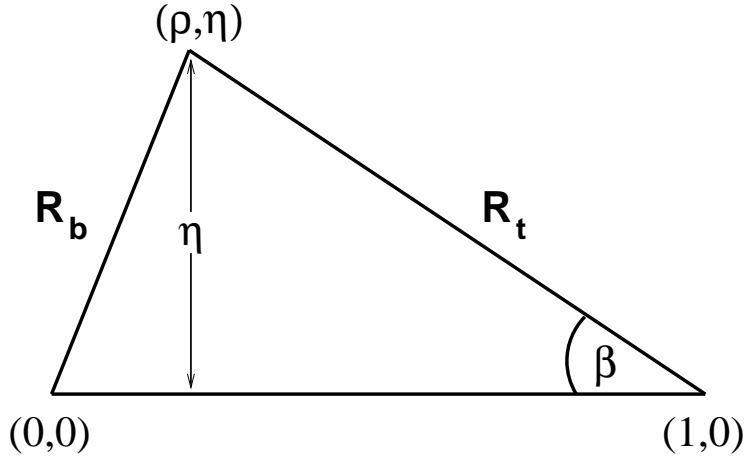}
     }}
    \caption[]{The ``true'' unitarity triangle.}
     \label{fig:unitarity}
\end{figure}

In the Standard Model, the present knowledge of the CKM matrix implies
$\varepsilon_1=+$ and $\varepsilon_2=+$. Furthermore $X_{\rm NL}$ and
$X(x_t)$ are real and the contribution proportional to $\IM \lambda_c
= -\IM \lambda_t$ can be safely neglected in view of $X_{\rm
  NL}/X(x_t) \sim O(10^{-3})$.  
Using the Wolfenstein parameterization \cite{WO} we have
\be\label{WP} 
\lambda_c=-\lambda, \quad\quad\quad
\lambda_t=-A^2\lambda^5 R_t e^{-i\beta} 
\ee 
and consequently
\be\label{F1} 
\RE F = -[ P_c+ A^2 R_t X(x_t) \cos\beta] 
\ee
and
\be
\label{F2} 
\IM F = A^2 R_t X(x_t) \sin\beta \, . 
\ee
Moreover one finds
\be\label{eta} 
\eta=R_t \sin \beta = \frac{\sqrt{B_2}}{A^2 X(x_t)}, 
\ee
where $\eta$ is the height of the unitarity triangle and $R_t$ is the
length of one of its sides as shown in
fig.~\ref{fig:unitarity}.  Consequently, the measurements of Br$(K^+
\to \pi^+ \nu \bar \nu )$ and Br$(K_L \to \pi^0 \nu \bar \nu )$ allow
the construction of the unitarity triangle.
Inserting (\ref{F1}) and (\ref{F2}) into (\ref{MAIN}) one derives
(\ref{cbb}) with $\sigma=1$. The corrections $\ord(\lambda^2)$ in
(\ref{cbb}) follow from the improvement \cite{BLO} of the Wolfenstein
parameterization. In order to simplify our presentation of the effects
of new physics we will neglect these corrections.  They can be
included in a straightforward manner if necessary. 

It should be stressed that (\ref{bkpn}), (\ref{bklpn}) and (\ref{MAIN})
are rather general as they are valid in any extension of the Standard
Model, provided
\begin{itemize}
\item
the new physics contributions to the tree level decay 
$K^+\to\pi^0 e^+\nu$ and
\item
the contributions of operators different from $\bar{s}_L \gamma^\mu
d_L \bar{\nu}^l_L \gamma_\mu \nu^l_L$ 
\end{itemize}
can be neglected.
Consequently, they are
useful for generalizations of this discussion to models in which
$F$ includes additional contributions not present in the Standard
Model. 

The neglect of new-physics contributions to
$K^+\to\pi^0 e^+\nu$ is very well justified in all known extensions of
the Standard Model. On the other hand, as pointed out in \cite{GN},
if lepton number is violated also  operators  $\bar{s}_L \gamma^\mu
d_L \bar{\nu}^i_L \gamma_\mu \nu^j_L$ 
 may contribute. In this case the
final state $\pi^0 \nu\bar\nu$ is not necessarily a CP eigenstate
and CP conserving contributions to $\klpn$ can in principle be
substantial. In our opinion such a situation is rather unlikely and
will not be considered in this paper.

\section{A Model Independent Analysis}
\label{sec:MI}

New physics can modify the effective Hamiltonian in (\ref{hkpn})
through new box diagram and penguin diagram contributions involving
new particles such as charged Higgs, charginos, stops etc. In the case
of $K_L \to\pi^0 \nu\bar\nu$ also new contributions to $K^0-\bar K^0$
mixing should in principle be considered. However, the smallness of
$\varepsilon_K=\ord(10^{-3})$ implies that the $K^0-\bar K^0$ mixing
effects should be safely negligible in $K_L \to\pi^0 \nu\bar\nu$
\cite{Littenberg} independent of the model considered as long as ${\rm
  Br}(\klpn)>\ord(10^{-13})$.  Since the Standard Model prediction
amounts to ${\rm Br}(\klpn)=(2.8\pm 1.7)\cdot 10^{-11}$ \cite{BF97} we
expect that this inequality is satisfied also in a large class of its
generalizations.  Consequently the only new physics contributions
relevant for $\klpn$ are the ones which modify the Hamiltonian in
(\ref{hkpn}).  Since this Hamiltonian governs both $\kpn$ and $\klpn$,
the discussion of new physics effects is considerably simplified
compared to the case in which the $K^0-\bar K^0$ mixing had to be
considered.  This situation is opposite to the one encountered in the
CP asymmetry $a_{\psi K_S}$, where one expects the dominant new
contributions from new phases in $B^0_d-\bar B^0_d$ mixing with
negligible contributions from new phases in the decay amplitude.

In order to simplify the analysis we will assume that:
\begin{itemize}
\item the unitarity of the three-generation CKM matrix is maintained;
\item the new-physics contributions in semi-leptonic tree-level
  $B$ decays used to determine $|V_{cb}|$ or $A$ as well as
  $|V_{ub}/V_{cb}|$ can be neglected.
\end{itemize} 
Then the new-physics contributions to $K\to\pi\nu\bar\nu$ can be
described quite generally by two new parameters $r_K$ and $\theta_K$
defined by
\be\label{new}
X_{\rm new}=r_K e^{-i\theta_K} X(x_t).
\ee
Here $X_{\rm new}$ summarizes all contributions to ${\cal H}_{\rm eff}$
except for the Standard Model contribution with internal charm
quarks. In a given model $r_K$ and $\theta_K$ are generally
complicated functions of masses of new particles and of new
couplings. They can also carry additional $m_t$ dependence resulting
from loop diagrams in which simultaneously the top quark and new
particles are exchanged. Well-known examples of such a situation
are box and Z-penguin diagrams with top and charged Higgs exchanges
present in two Higgs doublet models and  models
based on supersymmetry.

In principle we could also modify the first term in (\ref{hkpn}).
This, however, would unnecessarily complicate our discussion.  Any 
new-physics effects on the term proportional to $\lambda_c$ can be
included in the second term without the loss of generality. Yet one
has to remember that $\lambda_c\not=\lambda_t$ and any new physics
contribution proportional to $\lambda_c$ will not only modify $r_K$
but also give a contribution to $\theta_K$ even if this new
contribution does not carry any new phase. 

In this context, it should also be remarked that 
at least in the two Higgs doublet model the loop diagrams with charm
and charged Higgs exchanges give negligible contribution to
$\kpn$  because of very small charm-$H^\pm$ Yukawa couplings.

With $X(x_t)$ replaced by $X_{\rm new}$ we have
\be\label{F3}
\RE F = -[ P_c+ A^2 R_t r_K X(x_t) \cos(\beta+\theta_K)]
\ee
and
\be\label{F4}
\IM F= A^2 R_t r_K X(x_t) \sin(\beta+\theta_K)\, ,
\ee
which implies
\begin{equation}\label{nsin}
\sin 2(\beta+\theta_K)=\frac{2 r_s}{1+r^2_s},
\end{equation}
\begin{equation}\label{nbkpn}
{\rm Br}(\kpn)=\kappa_+\cdot\left[ P_c^2+A^4R^2_t r_K^2 X^2(x_t) 
           +2 P_c A^2 R_t r_K X(x_t) \cos (\beta+\theta_K)\right]
\end{equation}
and
\begin{equation}\label{nbklpn}
{\rm Br}(K_{\rm L}\to\pi^0\nu\bar\nu)=\kappa_{\rm L}\cdot 
A^4 R_t^2 r_K^2 X^2(x_t) \sin^2(\beta+\theta_K)\,,
\end{equation}
where $r_s$ is given by 
\be
r_s=  \frac{\varepsilon_1 \sqrt{B_1 - B_2} - P_c}{\varepsilon_2
  \sqrt{B_2}}.
\label{rss}
\ee
For $\varepsilon_1=+$, $\varepsilon_2=+$ and $\sigma=1$, $r_s$ reduces
to (\ref{cbb}). Finally, 
\be\label{etanew}
\varepsilon_2 \frac{\sqrt{B_2}}{A^2 X(x_t)} = 
R_t r_K \sin(\beta + \theta_K)\equiv
\eta_{\rm f}.
\ee
In order to simplify the discussion, we will assume that
$\varepsilon_1=+$ and $\varepsilon_2=+$ as in the Standard Model. 

\begin{figure}   
    \centerline{
    \epsfxsize=6truecm
    \rotate[r]{
    \epsffile{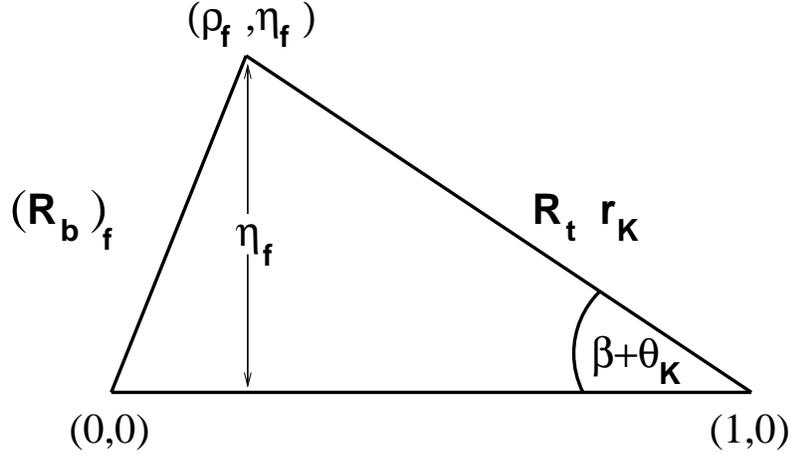}
     }}
    \caption[]{The ``fake'' unitarity triangle from $K \to \pi \nu
      \bar \nu$.}
     \label{fig:fakeunitarity}
\end{figure}

We make the following observations:
\begin{itemize}
\item If the new-physics contributions are governed by the CKM matrix
  and no contributions proportional to $\lambda_c$ as well as no new
  complex couplings are present $(\theta_K=0)$ then, even with
  $r_K\not=1$, $\sin 2\beta$ can be directly found by measuring ${\rm
    Br}(\kpn)$ and ${\rm Br}(\klpn)$ only. It does not require the
  knowledge of $r_K$ which depends on unknown masses of new particles
  and new couplings. Examples of such a situation are the two Higgs
  doublet model and the simplest supersymmetric models such as the
  ``constrained'' Minimal Supersymmetric Standard Model (MSSM), 
  to be specified in the following section.
\item Yet even in such situation the numerical value of $\beta$
  extracted using eq.~(\ref{nsin}) may generally differ from the value
  one would find using the Standard Model analysis of the unitarity
  triangle, which is based on $B^0_{d,s} - \bar{B}^0_{d,s}$ mixings
  and $\varepsilon_K$. Indeed, new-physics contributions to the latter
  quantities may require a modified value of $\beta$, which as we have
  seen can be directly extracted from eq.~(\ref{nsin}) if $\theta_K=0$.
\item In general, however, $\theta_K\not=0$ and as seen in
  (\ref{nsin}) ${\rm Br}(\kpn)$ and ${\rm Br}(\klpn)$ allow to extract only
  $(\beta+\theta_K)$ instead of $\beta$ as in the case of the Standard
  Model.  Once $(\beta+\theta_K)$ has been determined, one can use
  (\ref{nbkpn}) or (\ref{etanew}) to find the product $R_t r_K$.
\item We also observe that the ratio on the left-hand side of
  (\ref{etanew}) no longer measures $\eta$ as in (\ref{eta}) but a new
  quantity $\eta_{\rm f}$.
\item Thus in the presence of new physics the measurements of ${\rm
    Br}(\kpn)$ and ${\rm Br}(\klpn)$ imply a ``fake'' unitarity
  triangle shown in fig.~\ref{fig:fakeunitarity} in which $\beta$ and
  $\eta$ are replaced by $(\beta + \theta_K)$ and $\eta_{\rm f}$
  respectively.  This implies that the side $(R_b)_{\rm f}$ in this
  fake triangle will generally differ from $R_b$ defined as
  \be\label{Rb} 
  R_b=\frac{1}{\lambda}\left\vert
  \frac{V_{ub}}{V_{cb}}\right\vert.  
  \ee
\item In order to extract $\beta$, $\theta_K$, $R_t$ and $r_K$
  additional input is necessary.
\end{itemize}

Let us first remark that in order to see whether $\theta_K\not=0$ and
$r_K\not=1$ one can simply check whether $(R_b)_{\rm f}$ in
fig.~\ref{fig:fakeunitarity} agrees with $R_b$ in (\ref{Rb}) which is
expected to be unaffected by new-physics contributions. Similarly one
can use the 
standard analysis of the unitarity triangle \cite{BF97} which involves
the Standard Model expressions for $\varepsilon_K$, $B^0_d-\bar
B^0_d$ mixing and $|V_{ub}/V_{cb}|$. This analysis gives the values
of $\beta$ and $R_t$, albeit with substantial uncertainties.
Inserting these values into (\ref{nbkpn}) and (\ref{nbklpn}) and
comparing with the measured values of ${\rm Br}(\kpn)$ and ${\rm Br}(\klpn)$, may
give the first hint whether $r_K\not=1$ and $\theta_K\not=0$ are
indeed required.

Now the new physics will generally contribute also to $\varepsilon_K$
and $B^0_d-\bar B^0_d$ mixing. Consequently in order to extract the
values of $\theta_K$ and $r_K$ a more refined analysis is required.
To this end one can use the model independent construction of the
unitarity triangle of Grossman, Nir and Worah \cite{GNW}
 which involves the CP asymmetries $a_{\psi K_S}$ and $a_{\pi\pi}$ 
in $B_d\to\psi K_S$ and $B_d \to\pi^+\pi^-$ decays respectively.
This analysis is based on the following plausible assumption:
\begin{itemize}
\item
the $\bar b \to \bar c c \bar s$ and $\bar b \to \bar u ud$ decays for
$a_{\psi K_S}$ and $a_{\pi\pi}$ respectively are mediated by 
Standard Model tree level diagrams with negligible contributions 
from new physics in decay amplitudes. This implies that the only impact of
new physics on $a_{\psi K_S}$ and $a_{\pi\pi}$ will come through
$B^0_d-\bar B^0_d$ mixing.
\end{itemize}

The remaining three assumptions involving $K^0-\bar K^0$ mixing, the
dominance of the Standard Model contribution in the determination of
$|V_{ub}/V_{cb}|$ and the unitarity of the three generation CKM
matrix, made in \cite{GNW}, are the same as we have made in connection
with $K \to\pi\nu\bar\nu$.

A weak point of the analysis of \cite{GNW} is the use of the
asymmetry $a_{\pi\pi}$ to extract the angle $\alpha$.
It is well known that this extraction
is affected by the ``QCD penguin pollution''.
The recent CLEO results for penguin dominated decays indicate that
this pollution could be substantial \cite{ciuchini}.
The most popular strategy to deal with this ``penguin problem''
is the isospin analysis of Gronau and London \cite{CPASYM}. It
requires however the measurement of ${\rm Br}(B^0_d\to \pi^0\pi^0)$ which is
expected to be below $10^{-6}$: a very difficult experimental task.
For this reason several, rather involved, strategies \cite{SNYD} 
have been proposed which
avoid the use of $B^0_d \to \pi^0\pi^0$ in conjunction with
the asymmetry $a_{\pi\pi}$. 
 It is to be seen which of these methods
will eventually allow us to measure $\alpha$ with a respectable precision.
In what follows we will assume that all these problems will be overcome
and $\alpha$ will be measured through $a_{\pi\pi}$ one day.

Under these circumstances the new-physics contributions in the
analysis of \cite{GNW} can be described analogously to (\ref{new})
by two parameters $r_d$ and $\theta_d$, defined by
\be
\left(r_d e^{i\theta_d}\right)^2\equiv
\frac{\langle B^0_d|{\cal H}^{\rm full}_{\rm eff}(\Delta B=2)|
\bar B^0_d\rangle}
{\langle B^0_d|{\cal H}^{\rm SM}_{\rm eff}(\Delta B=2)|\bar B^0_d
\rangle}
\label{rdtd}
\ee
where ${\cal H}^{\rm full}_{\rm eff}(\Delta B=2)$ and 
${\cal H}^{\rm SM}_{\rm eff}(\Delta B=2)$
denote the full Hamiltonian (including the new and Standard Model
contributions) and the Standard Model Hamiltonian respectively.

With new-physics contributions to
$B^0_d-\bar B^0_d$ mixing the asymmetry $a_{\psi K_S}$ does no
longer measure $\sin 2\beta$ but $\sin 2(\beta+\theta_d)$ and the
relation (\ref{gr}) is not satisfied if $\theta_K\not=\theta_d$:
\be\label{grnew}
[\sin 2(\beta+\theta_K)]_{\pi\nu\bar\nu}\not=
[\sin 2(\beta+\theta_d)]_{\psi K_{\rm S}}
\ee
Since $\theta_K$ originates in new contributions to the decay
amplitude $K\to\pi\nu\bar\nu$ and $\theta_d$ in new contributions
to the $B^0_d-\bar B^0_d$ mixing, it is very likely that 
$\theta_K\not=\theta_d$.

Now as demonstrated in \cite{GNW}, the knowledge of
$a_{\psi K_S}$, $a_{\pi\pi}$ and $R_b$ allows to extract in a model
independent manner $\beta$, $R_t$ and $\theta_d$ and consequently the
true unitarity triangle. 
In order to find $r_d$ also the $B^0_d-\bar B^0_d$ mixing parameter $x_d$
subject to hadronic uncertainties has to be considered.
Inserting $R_t$ and
$\beta$ into (\ref{nsin})--(\ref{etanew}) allows then a model independent
extraction of $\theta_K$ and $r_K$ from $K\to\pi\nu\bar\nu$. This is
also evident from fig.~\ref{fig:fakeunitarity}.
This in turn gives more insight in the violation of the relation
(\ref{gr}) as anticipated in (\ref{grnew}).

Having extracted the values of $r_d$, $\theta_d$, $r_K$ and $\theta_K$
in a model independent manner, one can then compare these values with
explicit model calculations of these quantities. Such a comparison can
either exclude certain models or put bounds on their  parameters.

\begin{figure}   
    \centerline{
    \epsfxsize=6truecm
    \rotate[r]{
    \epsffile{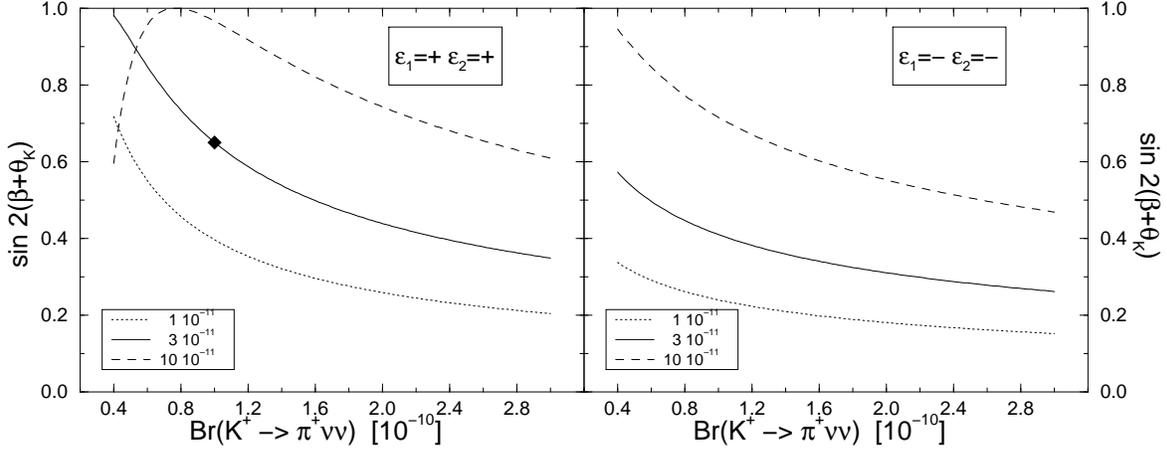}
     }}
    \caption[]{$\sin 2 (\beta + \theta_K)$ as a function of Br$(\kpn)$
      for Br$(\klpn)=1\cdot 10^{-11}$, $3 \cdot 10^{-11}$ and $1\cdot
      10^{-10}$ and two choices of
      $(\varepsilon_1,\varepsilon_2)$. The results for $(+,-)$ and
      $(-,+)$ are obtained by reversing the sign of $\sin 2 (\beta +
      \theta_K)$  in these plots.}
     \label{fig:ambiguities}
\end{figure}

Clearly in a general case one would have to take care of discrete
ambiguities represented by $\varepsilon_{1,2}$ in (\ref{rss}) and
(\ref{etanew}). Analogous discrete ambiguities are also present in the
analysis of ref. \cite{GNW}. In order to resolve these ambiguities,
additional experimental information will be needed.
In fig.~\ref{fig:ambiguities} we plot $\sin 2 (\beta +\theta_K)$ as a
function of Br(\kpnn) for different values of Br(\klpnn) and 
the choices $(+,+)$ and $(-,-)$ for
$(\varepsilon_1,\varepsilon_2)$. The results for $(+,-)$ and $(-,+)$
can be obtained by reversing the sign of $\sin 2 (\beta + \theta_K)$ in
the left and right plot in fig.~\ref{fig:ambiguities} respectively.
The black square is in the ball park of the central values expected in
the Standard Model. We observe that even a modest accuracy for  $\sin
2 (\beta + \theta_K)$ should be sufficient to distinguish between
various choices for $(\varepsilon_1,\varepsilon_2)$.

\section{$K \to \pi \nu \bar \nu$ and Supersymmetry}
\label{RS}

We will now study the effective Hamiltonian for $s \to d \nu \bar \nu$
transitions in the framework of a generalized SUSY extension of the
Standard Model with minimal field content and unbroken R-parity.

Supersymmetric contributions to the process $K^+ \to \pi^+ \nu \bar
\nu$ were already considered in refs.~\cite{bertolini1}-\cite{couture}
and, very recently, in ref.~\cite{nir}, which also analyzed $K_L \to
\pi^0 \nu \bar \nu$.  With the exception of refs.~\cite{nir,couture},
the phenomenological analyses have been done prior to the top quark
discovery and using much less stringent constraints on SUSY masses
than those presently available. This makes the comparison of our
numerical results with these papers very difficult. Moreover, most of
the previous studies were performed in the framework of the so-called
``constrained'' MSSM, in which
universality of the soft SUSY breaking terms is assumed. In this case,
all SUSY contributions to FCNC processes are still proportional to the
CKM mixing angles. Consequently, due to the heaviness of the
superpartners, they are generally expected to be small in comparison
with the Standard Model contributions.  Furthermore, in the case of
the ``constrained'' MSSM, the only non-negligible SUSY contribution is
proportional to $\lambda_t$, and therefore no new phase $\theta_K$ is
introduced in eq.~(\ref{new}). As already discussed in
sec.~\ref{sec:MI}, in this case it is still possible to extract $\sin
2 \beta$ from $K \to \pi \nu \bar \nu $ decays, since eq.~(\ref{sin})
remains unaffected.

However, it should be noted that the ``constrained'' MSSM is based on
very strong assumptions which do not find at present a strong
theoretical motivation. It is therefore useful to consider a more
general definition of the MSSM, in which the assumptions about the
universality of the soft SUSY breaking terms are relaxed. In this
``unconstrained'' version of the MSSM new sources of flavour and CP
violation are present in the mass matrices of sfermions, and in
general large contributions to FCNC and CP violating processes are
expected in this case. To be able to study the contributions to $s \to
d \nu \bar \nu$ transitions in such a generalized extension of the
Standard Model, and to take properly into account the constraints
coming from measured low-energy FCNC and CP-violating processes, one
needs some kind of a model-independent parameterization of the
flavour- and CP-violating quantities in SUSY. Such a parameterization
has been formulated in the framework of the mass-insertion
approximation \cite{hall}, and will be discussed in detail below.

\subsection{Computation of SUSY Contributions}
\label{Conto}
 
\begin{figure}   
    \begin{center}
\input{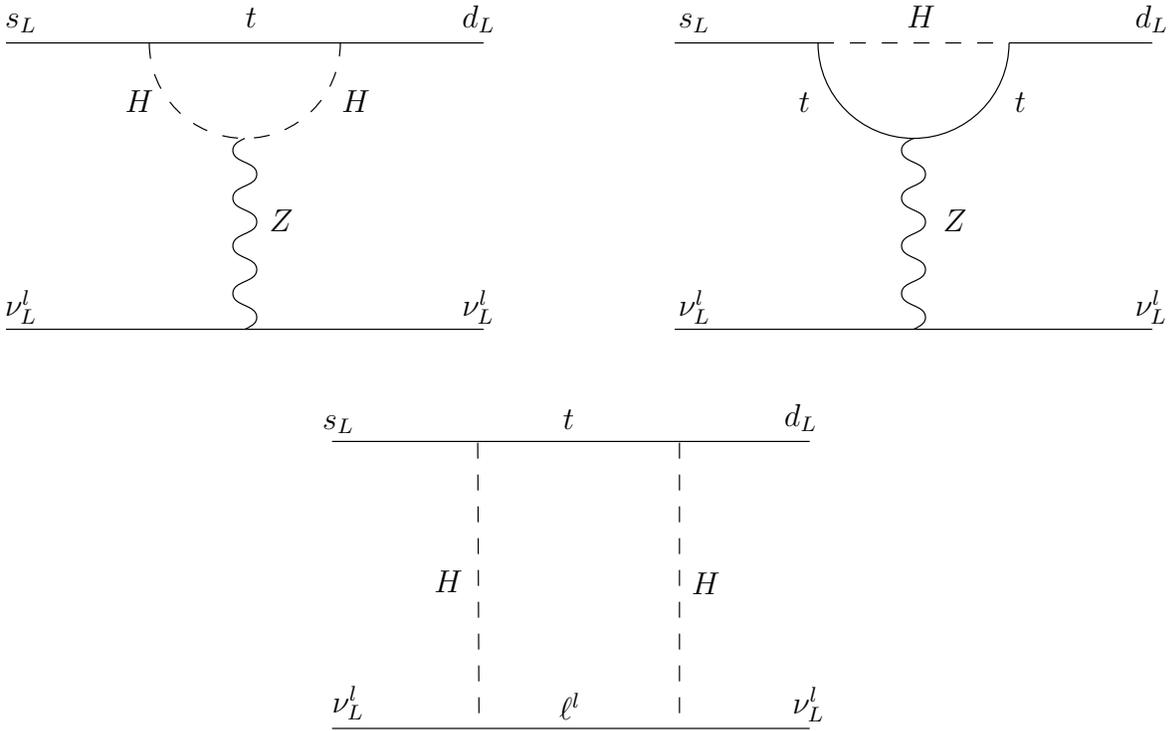}
    \end{center}
    \caption[]{Feynman diagrams for the charged-Higgs contribution to
      $s \to d \nu \bar \nu$ transitions.}
    \label{fig:diach}
\end{figure}

\begin{figure}   
    \begin{center}
\input{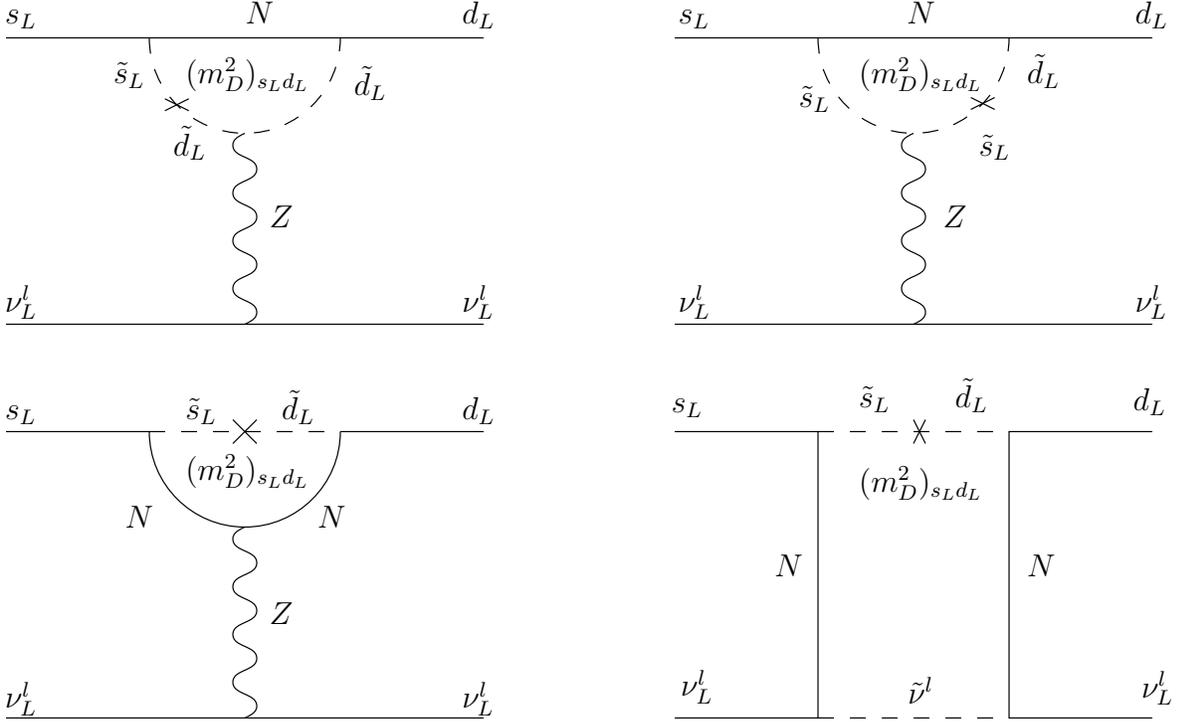}
    \end{center}
    \caption[]{Feynman diagrams for the neutralino contribution to
      $s \to d \nu \bar \nu$ transitions.}
    \label{fig:dian}
\end{figure}

There are several one-loop SUSY contributions to $s \to d \nu \bar
\nu$ transitions, according to the kind of diagram (penguin or box)
and to the virtual particles running in the loop: i) charged Higgses
and up-type quarks (see fig.~\ref{fig:diach}), ii) charginos and
up-type squarks (and charged sleptons in box diagrams, see
fig.~\ref{fig:diachi}), iii) gluinos and down-type squarks (see
fig.~\ref{fig:diag}), iv) neutralinos and down type squarks (and
sneutrinos in box diagrams, see fig.~\ref{fig:dian}).

\begin{figure}   
    \begin{center}
\input{diachi.pstex_t}
    \end{center}
    \caption[]{Feynman diagrams for the chargino contribution to
      $s \to d \nu \bar \nu$ transitions.}
    \label{fig:diachi}
\end{figure}
\begin{figure}   
    \begin{center}
\input{diag.pstex_t}
    \end{center}
    \caption[]{Feynman diagrams for the gluino contribution to
      $s \to d \nu \bar \nu$ transitions.}
    \label{fig:diag}
\end{figure}

First of all we note that, due to the left-handed chirality of the
neutrinos and due to Lorenz invariance, the quark fields entering in
dimension-six operators must have the same chirality. Therefore, there
are only two possible dimension-six local operators for $s \to d \nu
\bar \nu$ transitions: $\bar{s}_L \gamma_\mu d_L \bar{\nu}_L
\gamma^\mu \nu_L$ and $\bar{s}_R \gamma_\mu d_R \bar{\nu}_L \gamma^\mu
\nu_L$.  

Let us consider first gluino-mediated transitions. As already remarked
in ref.~\cite{bertolini}, these contributions are negligible.  First
of all, there are no gluino-mediated box diagrams. Concerning the
penguin diagrams, it is clear that if the $Z$ coupling to squarks were
proportional to the photon one, the amplitude would be suppressed at
least by a factor $q/M_Z$ with respect to the Standard Model one,
since by gauge invariance there is no contribution to the effective $s
d \gamma$ vertex of zeroth order in the external momenta (we denote by
$q$ the momentum of the $Z$ boson).  In fact, gauge invariance
constrains the effective $s d \gamma$ vertex to be of the form $F_1
\bar{s}_L \left( q^2 \gamma_\mu - q_\mu \qslash \right) d_L + F_2
\bar{s}_R \sigma_{\mu\nu} q^\nu d_L$.  This is indeed the case if the
squarks running in the loop have fixed ``chirality''. To avoid the
$q/M_Z$ suppression, one has to perform a double chirality flip in the
down-type squark propagator, which is once again highly suppressed
under the plausible assumption that the left-right mixing of squarks
of the first two generations is negligible\footnote{This is also the
  reason why the $\sin^2 \Theta_W$ part of the $Z$ coupling does not
  contribute to the total amplitude.}.

Let us consider next the operators that the contributions in
figs.~\ref{fig:diach}, \ref{fig:diachi} and \ref{fig:dian}
could generate in the effective theory. ``RL'' operators as $\bar{s}_R
\sigma_{\mu\nu} q^\nu d_L \bar{\nu}_L \gamma^\mu \nu_L$ are suppressed
by $q/M_Z$ and either by a quark mass or a chirality flip in the
down-type squark propagator. ``RR'' operators as $\bar{s}_R \gamma_\mu
d_R \bar{\nu}_L \gamma^\mu \nu_L$ can only be generated by a
neutralino exchange through U(1) gauge couplings and turn out to be
negligible. We are left with ``LL'' operators as $\bar{s}_L \gamma_\mu
d_L \bar{\nu}_L \gamma^\mu \nu_L$. In a general supersymmetric
extension of the Standard Model lepton flavour violation can be
present due to a misalignment between lepton and slepton mass
eigenstates.  Therefore operator involving different neutrino flavours
can be generated. However, the experimental limits on $\mu \rightarrow
e \gamma$, $\tau \rightarrow \mu \gamma$ and other lepton flavour
violating processes make their contribution to $K \rightarrow
\pi\bar{\nu}\nu$ negligible. The operators we will consider are
therefore the same as in the Standard Model. As in that case, we could
have a dependence of the coefficient on the neutrino flavour involved,
here due to a possible non-degeneracy of the sleptons running in the
boxes. For our purposes, it is sufficient to use a ``mean'' value for the
slepton masses of a given charge, without distinguishing among the three
neutrinos, as done for the top contribution in the Standard Model.

Our effective Hamiltonian for $K \rightarrow \pi\bar{\nu}\nu$ is
therefore given by eq.~(\ref{hkpn}) with $X(x_t)$ replaced by
\begin{equation}
X_{\rm new} = X(x_t) + X_H(x_{tH}) +C_\chi + C_N \, ,
\label{hesusy}
\end{equation}
where $x_{tH}=m_t^2/m_{H^\pm}^2$ and the function
$X_H(x_{tH})$ corresponding to the charged Higgs contribution is given in
the Appendix. $C_\chi$ and $C_N$ denote respectively the chargino and
neutralino contributions.

While the flavour structure of the Standard Model and charged Higgs
contributions involves just the CKM matrix, $C_N$ and $C_\chi$ depend
on the transition matrices between quark and squark mass eigenstates,
whose determination requires the diagonalization of the $6\times 6$
squark mass matrices. In the so-called super-CKM basis, in which the 
flavour structure
of the quark--squark--gaugino vertices is the same as in the 
quark--quark--gauge boson vertices, the squark mass matrices are given
by:
\begin{equation}
M^2_{D} = 
\left(\begin{array}{cc}
({\bf m}^2_D)_{LL} & ({\bf m}^2_D)_{LR} \\
({\bf m}^2_D)_{LR}^\dagger & ({\bf m}^2_D)_{RR}
\end{array}\right)
\qquad
M^2_{U} = 
\left(\begin{array}{cc}
({\bf m}^2_U)_{LL} & ({\bf m}^2_U)_{LR} \\
({\bf m}^2_U)_{LR}^\dagger & ({\bf m}^2_U)_{RR}
\end{array}\right)\, ,
\label{massesCKM}
\end{equation}
where
\begin{eqnarray}
        ({\bf m}^{2}_U)_{LL} &=&
        {\bf m}^{2}_{\tilde{u}_{L}} + \left({\bf m}^{u}\right)^2 +
        \frac{M_{Z}^{2}}{6}\left(3-4\sin^{2} \Theta_W\right)
        \cos 2\beta \,{\bf 1}
        \nonumber \\
        ({\bf m}^{2}_U)_{LR} &=&
        -\mu {\bf m}^{u} \cot \beta -\frac{v \sin \beta}{\sqrt{2}}
        {\bf A}^{u}
        \nonumber \\
        ({\bf m}^{2}_U)_{RR} &=&
        {\bf m}^{2}_{\tilde{u}_{R}} +
        \left({\bf m}^{u}\right)^2 + 
        \frac{2}{3}M_{Z}^{2}\sin^{2}\Theta_W \cos 2 \beta\, {\bf 1}
        \nonumber \\
        ({\bf m}^{2}_D)_{LL} &=&
        {\bf m}^{2}_{\tilde{d}_{L}} +
        \left({\bf m}^{d}\right)^2 -
        \frac{M_{Z}^{2}}{6}\left(3-2\sin^{2} \Theta_W\right)\cos
        2\beta\, 
        {\bf 1}
        \nonumber \\
        ({\bf m}^{2}_D)_{LR} &=&
        -\mu {\bf m}^{d} \tan \beta -\frac{v \cos \beta}{\sqrt{2}}
        {\bf A}^{d}
        \nonumber \\
        ({\bf m}^{2}_D)_{RR} &=&
        {\bf m}^{2}_{\tilde{d}_{R}} +
        \left({\bf m}^{d}\right)^2 -
        \frac{1}{3}M_{Z}^{2}\sin^{2}\Theta_W 
        \cos 2 \beta \,{\bf 1}.
        \label{sqmass2}
\end{eqnarray}
We denote by ${\bf m}^{2}_{\tilde{u}_{L}}$, ${\bf
  m}^{2}_{\tilde{u}_{R}}$,  ${\bf m}^{2}_{\tilde{d}_{L}}$ and 
${\bf m}^{2}_{\tilde{d}_{R}}$ the soft-breaking masses for squarks and
by ${\bf A}^{u}$ and ${\bf A}^{d}$ the soft-breaking trilinear
couplings for up and down squarks in this basis, at the electroweak scale.
The quark mass matrices are diagonal in this basis: 
\begin{eqnarray}
        {\bf m}^{u} & = & \diag \left(m_{u},m_{c},m_t\right),
        \nonumber \\
        {\bf m}^{d} & = & \diag \left(m_{d},m_{s},m_b\right).
        \label{dmass}
\end{eqnarray}

In order to obtain simple analytic formulae in a generic model, it is
useful not to perform the exact diagonalization of the squark mass
matrices, but instead use the mass-insertion approximation. Since we
are interested in $s\to d$ transitions, we switch to a basis in which
the $d^i_L-\tilde{d}^j_L-N_n$ and $d^i_L-\tilde{u}^j_L-\chi_n$
couplings are flavour diagonal, and the flavour change in the
left-handed sector is exhibited by the non-diagonality of the sfermion
propagators.  Denoting by $\Delta$ the off-diagonal terms in the
sfermion mass matrices, the sfermion propagators can be expanded as a
series in terms of $\delta = \Delta/ \tilde{m}^2$, where $\tilde{m}$
is the average sfermion mass. As long as $\Delta$ is significantly
smaller than $\tilde{m}^2$, we can just take the first term of this
expansion and compute any given process in terms of these
$\delta$'s\footnote{Care must be taken in the truncation of the
  expansion at the first order in those models where there is a strong
  hierarchy between different mass insertions. For example, in the
  constrained MSSM, in the notations of eq.~(\ref{massesCKM}),
  $(m^2_D)_{d_L s_L}/\tilde{m}^2 \sim (m^2_D)_{d_L b_L} (m^2_D)_{b_L
    s_L}/\tilde{m}^4$.}. This is equivalent to a first order
diagonalization of the squark mass matrices around their diagonal
part.

The basis we have chosen is obtained starting from the super-CKM basis
by rotating the left-handed up-type squark fields:
\be
\tilde{u}_L \longrightarrow V^\dagger \tilde{u}_L\,,
\label{transform}
\ee
where $V$ is the CKM matrix. In our basis, the up-squark mass matrix is
given in terms of eqs.~(\ref{massesCKM}) and (\ref{sqmass2}) by
\begin{equation}
{\cal M}^2_U = 
\left(\begin{array}{cc}
({\bf m}^2_U)_{d_L d_L} & ({\bf m}^2_U)_{d_L u_R} \\
({\bf m}^2_U)_{u_R d_L} & ({\bf m}^2_U)_{u_R u_R}
\end{array}\right),
\label{massesrom}
\end{equation}
with the notation
\begin{eqnarray}
({\bf m}^2_U)_{d_L d_L} = V^\dagger ({\bf m}^2_U)_{LL} V & \quad &
({\bf m}^2_U)_{d_L u_R} = V^\dagger ({\bf m}^2_U)_{LR} \nonumber \\
({\bf m}^2_U)_{u_R d_L} = ({\bf m}^2_U)_{RL} V & \quad &
({\bf m}^2_U)_{u_R u_R} = ({\bf m}^2_U)_{RR}.
\label{massesrom2}
\end{eqnarray}
In the above formulae, $u$ and $d$
represent the 
three-generation space: $u$ stands for $u$, $c$, $t$ and analogously
for $d$. 
 
In principle one should diagonalize exactly the up-squark mass matrix
in the $\tilde{t}_L-\tilde{t}_R$ sector before performing the
mass-insertion expansion. However, this would make the equations more
cumbersome without adding any new feature in the results, and
therefore we will stick to the basis above. It is a straightforward
exercise to include this exact diagonalization in our formulae.

The chargino $(C_\chi)$ and neutralino $(C_N)$ contributions are
given by
\begin{eqnarray}
C_\chi &=& X_\chi^0 +
X_\chi^{LL} R^U_{s_L d_L} + X_\chi^{LR} R^U_{s_L t_R} + {X_\chi^{LR}}^*
R^U_{t_R d_L},
\nonumber \\
C_N &=& X_N R^D_{s_L d_L},
\label{CSUSY}
\end{eqnarray}
where the functions $X_\chi^i$ and $X_N$, which depend on SUSY masses
and respectively on chargino and neutralino mixing angles, are given
in the Appendix. The $R$ parameters are defined in terms of the mass
insertions in the following way:
\begin{eqnarray}
\label{normalization}
R^D_{s_L d_L} = \frac{(m^2_D)_{s_L d_L}}{V_{ts}^*
 V_{td}m^2_{\tilde{d}_L}} & \qquad & 
R^U_{s_L d_L} = \frac{(m^2_U)_{s_L d_L}}{V_{ts}^*
 V_{td}m^2_{\tilde{u}_L}} \nonumber \\
R^U_{s_L t_R} = \frac{(m^2_U)_{s_L t_R}}{V_{ts}^*
m_{\tilde{u}_L} m_t} & \qquad & 
R^U_{t_R d_L} = \frac{(m^2_U)_{t_R d_L}}{
 V_{td}m_t m_{\tilde{u}_L}},
\end{eqnarray}
where $m^2_{\tilde{d}_L}$ and $m^2_{\tilde{u}_L}$ are respectively 
the squared
masses of the down- and up-type squarks of the first two generations.

Eq.~(\ref{CSUSY}) shows explicitly the dependence of the various
contributions on the off-diagonal entries of the squark mass matrices,
that are completely unknown apart from upper limits on their real and
imaginary parts, and are strongly model dependent. The dependence on
masses and other supersymmetric parameters is on the other hand
contained in the $X$ functions, whose order of magnitude is rather
model independent and can be read from a scatter plot for random
values of the relevant SUSY parameters (see below).

The model dependence is therefore in this way almost completely
contained in the $R$ parameters. We have chosen the normalization of
the $R$'s in such a manner that their absolute value in the
constrained MSSM is expected to be of order one or
smaller. Indeed, a rough estimate gives
\begin{equation}
\label{RMSSM1}
(R^D_{s_L d_L})_{\rm MSSM} \sim (R^U_{s_L d_L})_{\rm MSSM} \sim
-\frac{3}{(4 \pi)^2}  Y^2_t \log \frac{M^2_0}{M_Z^2}
\end{equation}
and
\begin{equation}
\label{RMSSM2}
(R^U_{s_L t_R})_{\rm MSSM} \sim (R^U_{t_R d_L})_{\rm MSSM} \sim
-\frac{A_t + \mu \cot \beta}{m_{\tilde{u}_L}},
\end{equation}
where $Y_t$ is the top Yukawa coupling, $Y_t A_t$ is the (3,3) element
of the trilinear
scalar coupling in the
up sector and $M_0$ is the universality scale. For $M_0 = {\cal
O}(M_{\rm GUT})$, $\frac{3}{(4 \pi)^2} \log \frac{M^2_0}{M_Z^2} Y^2_t
= {\cal O}(1)$ and therefore $(R^D_{s_L d_L})_{\rm MSSM} \sim (R^U_{s_L
d_L})_{\rm MSSM} \sim 1$. 
Experimental limits on the
$R$ parameters are discussed below and summarized in
table~\ref{table}.

\begin{table}
 \begin{center}
 \begin{tabular}{||c|c||}  \hline \hline
 quantity & upper limit\\
 \hline
 & \\
 $R^D_{s_L d_L}$ & $(-112 - 55 \, i)\,\frac{m_{\tilde{d}_L}}{500 \,
   {\rm GeV}}$\\
 & \\
 $R^U_{s_L d_L}$ & $(-112 - 54 \, i)\,\frac{m_{\tilde{u}_L}}{500 \,
   {\rm GeV}}$\\
 & \\
 $R^U_{s_L t_R}$ &
 ${\rm Min} \left\{231\,\left(\frac{m_{\tilde{u}_L}}{500 \, {\rm GeV}}
 \right)^3\,,43\right\}\times e^{i\phi}$, $0<\phi<2 \pi$\\
 & \\
 $R^U_{t_R d_L}$ &  $37\,\left(\frac{m_{\tilde{u}_L}}{500 \, {\rm
       GeV}}\right)^2 \times e^{i \phi}$, $0<\phi<2 \pi$\\
 & \\
 \hline \hline
 \end{tabular}
 \caption[]{Upper limits for the $R$ parameters. Notice that the phase
   of $R^U_{s_L t_R}$ and $R^U_{t_R d_L}$ is unconstrained. See
   Section \ref{sec:genres} for details on how these upper bounds are
   derived.}
 \label{table}
 \end{center}
\end{table} 

\subsection{Phenomenological analysis}
\label{pheno}

We are now ready to write a formula for $r_K e^{-i\theta_K}$:
\begin{equation}
\label{r}
r_K e^{-i\theta_K} = 1 + \frac{X_H}{X} + \frac{X_\chi^0}{X} +
\frac{X_\chi^{LL}}{X} R^U_{s_L d_L} + \frac{X_\chi^{LR}}{X} 
R^U_{s_L t_R} + \frac{{X_\chi^{LR}}^*}{X} R^U_{t_R d_L} + 
\frac{X_N}{X} R^D_{s_L d_L}
\end{equation}
where $X\equiv X(x_t)$. The amount of SUSY corrections to $K \to \pi
\nu \bar \nu$ branching ratios and $\sin 2 \beta$ depends on how much
$r_K e^{-i\theta_K}$ deviates from unity. This depends of course on the
size of the mass insertions, that is largely unknown. 

\subsubsection{Results in the MSSM}
\label{sec:MSSMres}

\begin{figure}   
    \begin{center}
       \input{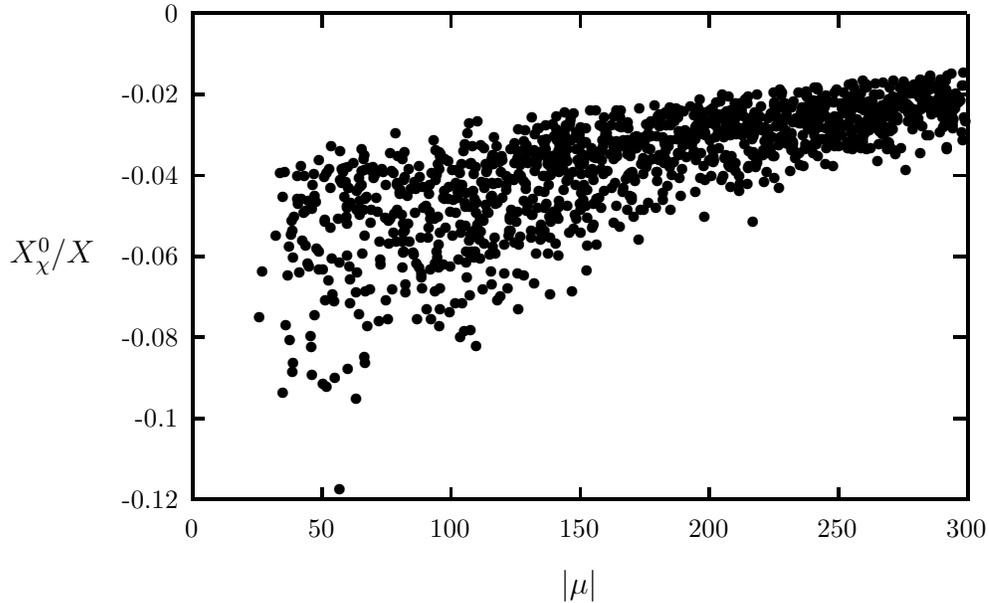}
    \end{center}
    \caption[]{The ratio $X_\chi^0/X$ as a function of $\vert \mu \vert$.}
     \label{fig:0}
\end{figure}

\begin{figure}   
    \begin{center}
\setlength{\unitlength}{0.240900pt}
\ifx\plotpoint\undefined\newsavebox{\plotpoint}\fi
\begin{picture}(1500,900)(0,0)
\font\gnuplot=cmr10 at 10pt
\gnuplot
\sbox{\plotpoint}{\rule[-0.500pt]{1.000pt}{1.000pt}}%
\put(220.0,113.0){\rule[-0.500pt]{4.818pt}{1.000pt}}
\put(198,113){\makebox(0,0)[r]{0.005}}
\put(1416.0,113.0){\rule[-0.500pt]{4.818pt}{1.000pt}}
\put(220.0,222.0){\rule[-0.500pt]{4.818pt}{1.000pt}}
\put(198,222){\makebox(0,0)[r]{0.01}}
\put(1416.0,222.0){\rule[-0.500pt]{4.818pt}{1.000pt}}
\put(220.0,331.0){\rule[-0.500pt]{4.818pt}{1.000pt}}
\put(198,331){\makebox(0,0)[r]{0.015}}
\put(1416.0,331.0){\rule[-0.500pt]{4.818pt}{1.000pt}}
\put(220.0,440.0){\rule[-0.500pt]{4.818pt}{1.000pt}}
\put(198,440){\makebox(0,0)[r]{0.02}}
\put(1416.0,440.0){\rule[-0.500pt]{4.818pt}{1.000pt}}
\put(220.0,550.0){\rule[-0.500pt]{4.818pt}{1.000pt}}
\put(198,550){\makebox(0,0)[r]{0.025}}
\put(1416.0,550.0){\rule[-0.500pt]{4.818pt}{1.000pt}}
\put(220.0,659.0){\rule[-0.500pt]{4.818pt}{1.000pt}}
\put(198,659){\makebox(0,0)[r]{0.03}}
\put(1416.0,659.0){\rule[-0.500pt]{4.818pt}{1.000pt}}
\put(220.0,768.0){\rule[-0.500pt]{4.818pt}{1.000pt}}
\put(198,768){\makebox(0,0)[r]{0.035}}
\put(1416.0,768.0){\rule[-0.500pt]{4.818pt}{1.000pt}}
\put(220.0,877.0){\rule[-0.500pt]{4.818pt}{1.000pt}}
\put(198,877){\makebox(0,0)[r]{0.04}}
\put(1416.0,877.0){\rule[-0.500pt]{4.818pt}{1.000pt}}
\put(220.0,113.0){\rule[-0.500pt]{1.000pt}{4.818pt}}
\put(220,68){\makebox(0,0){200}}
\put(220.0,857.0){\rule[-0.500pt]{1.000pt}{4.818pt}}
\put(372.0,113.0){\rule[-0.500pt]{1.000pt}{4.818pt}}
\put(372,68){\makebox(0,0){300}}
\put(372.0,857.0){\rule[-0.500pt]{1.000pt}{4.818pt}}
\put(524.0,113.0){\rule[-0.500pt]{1.000pt}{4.818pt}}
\put(524,68){\makebox(0,0){400}}
\put(524.0,857.0){\rule[-0.500pt]{1.000pt}{4.818pt}}
\put(676.0,113.0){\rule[-0.500pt]{1.000pt}{4.818pt}}
\put(676,68){\makebox(0,0){500}}
\put(676.0,857.0){\rule[-0.500pt]{1.000pt}{4.818pt}}
\put(828.0,113.0){\rule[-0.500pt]{1.000pt}{4.818pt}}
\put(828,68){\makebox(0,0){600}}
\put(828.0,857.0){\rule[-0.500pt]{1.000pt}{4.818pt}}
\put(980.0,113.0){\rule[-0.500pt]{1.000pt}{4.818pt}}
\put(980,68){\makebox(0,0){700}}
\put(980.0,857.0){\rule[-0.500pt]{1.000pt}{4.818pt}}
\put(1132.0,113.0){\rule[-0.500pt]{1.000pt}{4.818pt}}
\put(1132,68){\makebox(0,0){800}}
\put(1132.0,857.0){\rule[-0.500pt]{1.000pt}{4.818pt}}
\put(1284.0,113.0){\rule[-0.500pt]{1.000pt}{4.818pt}}
\put(1284,68){\makebox(0,0){900}}
\put(1284.0,857.0){\rule[-0.500pt]{1.000pt}{4.818pt}}
\put(1436.0,113.0){\rule[-0.500pt]{1.000pt}{4.818pt}}
\put(1436,68){\makebox(0,0){1000}}
\put(1436.0,857.0){\rule[-0.500pt]{1.000pt}{4.818pt}}
\put(220.0,113.0){\rule[-0.500pt]{292.934pt}{1.000pt}}
\put(1436.0,113.0){\rule[-0.500pt]{1.000pt}{184.048pt}}
\put(220.0,877.0){\rule[-0.500pt]{292.934pt}{1.000pt}}
\put(1,495){\makebox(0,0){$X_H/X$}}
\put(784,-22){\makebox(0,0){$m_H$ (GeV)}}
\put(220.0,113.0){\rule[-0.500pt]{1.000pt}{184.048pt}}
\put(244,790){\usebox{\plotpoint}}
\multiput(245.83,781.66)(0.496,-0.870){42}{\rule{0.120pt}{2.010pt}}
\multiput(241.92,785.83)(25.000,-39.828){2}{\rule{1.000pt}{1.005pt}}
\multiput(270.83,738.04)(0.496,-0.822){40}{\rule{0.120pt}{1.917pt}}
\multiput(266.92,742.02)(24.000,-36.022){2}{\rule{1.000pt}{0.958pt}}
\multiput(294.83,698.56)(0.496,-0.758){40}{\rule{0.120pt}{1.792pt}}
\multiput(290.92,702.28)(24.000,-33.281){2}{\rule{1.000pt}{0.896pt}}
\multiput(318.83,662.32)(0.496,-0.666){42}{\rule{0.120pt}{1.610pt}}
\multiput(314.92,665.66)(25.000,-30.658){2}{\rule{1.000pt}{0.805pt}}
\multiput(343.83,628.60)(0.496,-0.631){40}{\rule{0.120pt}{1.542pt}}
\multiput(339.92,631.80)(24.000,-27.800){2}{\rule{1.000pt}{0.771pt}}
\multiput(367.83,597.95)(0.496,-0.588){40}{\rule{0.120pt}{1.458pt}}
\multiput(363.92,600.97)(24.000,-25.973){2}{\rule{1.000pt}{0.729pt}}
\multiput(391.83,569.48)(0.496,-0.524){42}{\rule{0.120pt}{1.330pt}}
\multiput(387.92,572.24)(25.000,-24.240){2}{\rule{1.000pt}{0.665pt}}
\multiput(416.83,542.64)(0.496,-0.504){40}{\rule{0.120pt}{1.292pt}}
\multiput(412.92,545.32)(24.000,-22.319){2}{\rule{1.000pt}{0.646pt}}
\multiput(439.00,520.68)(0.504,-0.496){38}{\rule{1.293pt}{0.119pt}}
\multiput(439.00,520.92)(21.315,-23.000){2}{\rule{0.647pt}{1.000pt}}
\multiput(463.00,497.68)(0.550,-0.496){36}{\rule{1.386pt}{0.119pt}}
\multiput(463.00,497.92)(22.123,-22.000){2}{\rule{0.693pt}{1.000pt}}
\multiput(488.00,475.68)(0.580,-0.495){32}{\rule{1.450pt}{0.119pt}}
\multiput(488.00,475.92)(20.990,-20.000){2}{\rule{0.725pt}{1.000pt}}
\multiput(512.00,455.68)(0.611,-0.495){30}{\rule{1.513pt}{0.119pt}}
\multiput(512.00,455.92)(20.859,-19.000){2}{\rule{0.757pt}{1.000pt}}
\multiput(536.00,436.68)(0.684,-0.494){26}{\rule{1.662pt}{0.119pt}}
\multiput(536.00,436.92)(20.551,-17.000){2}{\rule{0.831pt}{1.000pt}}
\multiput(560.00,419.68)(0.714,-0.494){26}{\rule{1.721pt}{0.119pt}}
\multiput(560.00,419.92)(21.429,-17.000){2}{\rule{0.860pt}{1.000pt}}
\multiput(585.00,402.68)(0.777,-0.493){22}{\rule{1.850pt}{0.119pt}}
\multiput(585.00,402.92)(20.160,-15.000){2}{\rule{0.925pt}{1.000pt}}
\multiput(609.00,387.68)(0.777,-0.493){22}{\rule{1.850pt}{0.119pt}}
\multiput(609.00,387.92)(20.160,-15.000){2}{\rule{0.925pt}{1.000pt}}
\multiput(633.00,372.68)(0.871,-0.492){20}{\rule{2.036pt}{0.119pt}}
\multiput(633.00,372.92)(20.775,-14.000){2}{\rule{1.018pt}{1.000pt}}
\multiput(658.00,358.68)(0.977,-0.491){16}{\rule{2.250pt}{0.118pt}}
\multiput(658.00,358.92)(19.330,-12.000){2}{\rule{1.125pt}{1.000pt}}
\multiput(682.00,346.68)(0.900,-0.492){18}{\rule{2.096pt}{0.118pt}}
\multiput(682.00,346.92)(19.649,-13.000){2}{\rule{1.048pt}{1.000pt}}
\multiput(706.00,333.68)(1.118,-0.489){14}{\rule{2.523pt}{0.118pt}}
\multiput(706.00,333.92)(19.764,-11.000){2}{\rule{1.261pt}{1.000pt}}
\multiput(731.00,322.68)(1.070,-0.489){14}{\rule{2.432pt}{0.118pt}}
\multiput(731.00,322.92)(18.953,-11.000){2}{\rule{1.216pt}{1.000pt}}
\multiput(755.00,311.68)(1.182,-0.487){12}{\rule{2.650pt}{0.117pt}}
\multiput(755.00,311.92)(18.500,-10.000){2}{\rule{1.325pt}{1.000pt}}
\multiput(779.00,301.68)(1.235,-0.487){12}{\rule{2.750pt}{0.117pt}}
\multiput(779.00,301.92)(19.292,-10.000){2}{\rule{1.375pt}{1.000pt}}
\multiput(804.00,291.68)(1.182,-0.487){12}{\rule{2.650pt}{0.117pt}}
\multiput(804.00,291.92)(18.500,-10.000){2}{\rule{1.325pt}{1.000pt}}
\multiput(828.00,281.68)(1.501,-0.481){8}{\rule{3.250pt}{0.116pt}}
\multiput(828.00,281.92)(17.254,-8.000){2}{\rule{1.625pt}{1.000pt}}
\multiput(852.00,273.68)(1.381,-0.485){10}{\rule{3.028pt}{0.117pt}}
\multiput(852.00,273.92)(18.716,-9.000){2}{\rule{1.514pt}{1.000pt}}
\multiput(877.00,264.68)(1.501,-0.481){8}{\rule{3.250pt}{0.116pt}}
\multiput(877.00,264.92)(17.254,-8.000){2}{\rule{1.625pt}{1.000pt}}
\multiput(901.00,256.69)(1.746,-0.475){6}{\rule{3.679pt}{0.114pt}}
\multiput(901.00,256.92)(16.365,-7.000){2}{\rule{1.839pt}{1.000pt}}
\multiput(925.00,249.68)(1.570,-0.481){8}{\rule{3.375pt}{0.116pt}}
\multiput(925.00,249.92)(17.995,-8.000){2}{\rule{1.688pt}{1.000pt}}
\multiput(950.00,241.69)(1.746,-0.475){6}{\rule{3.679pt}{0.114pt}}
\multiput(950.00,241.92)(16.365,-7.000){2}{\rule{1.839pt}{1.000pt}}
\multiput(974.00,234.69)(2.119,-0.462){4}{\rule{4.250pt}{0.111pt}}
\multiput(974.00,234.92)(15.179,-6.000){2}{\rule{2.125pt}{1.000pt}}
\multiput(998.00,228.69)(1.827,-0.475){6}{\rule{3.821pt}{0.114pt}}
\multiput(998.00,228.92)(17.068,-7.000){2}{\rule{1.911pt}{1.000pt}}
\multiput(1023.00,221.69)(2.119,-0.462){4}{\rule{4.250pt}{0.111pt}}
\multiput(1023.00,221.92)(15.179,-6.000){2}{\rule{2.125pt}{1.000pt}}
\multiput(1047.00,215.71)(3.037,-0.424){2}{\rule{5.050pt}{0.102pt}}
\multiput(1047.00,215.92)(13.518,-5.000){2}{\rule{2.525pt}{1.000pt}}
\multiput(1071.00,210.69)(2.222,-0.462){4}{\rule{4.417pt}{0.111pt}}
\multiput(1071.00,210.92)(15.833,-6.000){2}{\rule{2.208pt}{1.000pt}}
\multiput(1096.00,204.71)(3.037,-0.424){2}{\rule{5.050pt}{0.102pt}}
\multiput(1096.00,204.92)(13.518,-5.000){2}{\rule{2.525pt}{1.000pt}}
\multiput(1120.00,199.71)(3.037,-0.424){2}{\rule{5.050pt}{0.102pt}}
\multiput(1120.00,199.92)(13.518,-5.000){2}{\rule{2.525pt}{1.000pt}}
\multiput(1144.00,194.71)(3.037,-0.424){2}{\rule{5.050pt}{0.102pt}}
\multiput(1144.00,194.92)(13.518,-5.000){2}{\rule{2.525pt}{1.000pt}}
\multiput(1168.00,189.71)(3.207,-0.424){2}{\rule{5.250pt}{0.102pt}}
\multiput(1168.00,189.92)(14.103,-5.000){2}{\rule{2.625pt}{1.000pt}}
\multiput(1193.00,184.71)(3.037,-0.424){2}{\rule{5.050pt}{0.102pt}}
\multiput(1193.00,184.92)(13.518,-5.000){2}{\rule{2.525pt}{1.000pt}}
\put(1217,177.92){\rule{5.782pt}{1.000pt}}
\multiput(1217.00,179.92)(12.000,-4.000){2}{\rule{2.891pt}{1.000pt}}
\put(1241,173.92){\rule{6.023pt}{1.000pt}}
\multiput(1241.00,175.92)(12.500,-4.000){2}{\rule{3.011pt}{1.000pt}}
\put(1266,169.92){\rule{5.782pt}{1.000pt}}
\multiput(1266.00,171.92)(12.000,-4.000){2}{\rule{2.891pt}{1.000pt}}
\put(1290,165.92){\rule{5.782pt}{1.000pt}}
\multiput(1290.00,167.92)(12.000,-4.000){2}{\rule{2.891pt}{1.000pt}}
\put(1314,161.92){\rule{6.023pt}{1.000pt}}
\multiput(1314.00,163.92)(12.500,-4.000){2}{\rule{3.011pt}{1.000pt}}
\put(1339,158.42){\rule{5.782pt}{1.000pt}}
\multiput(1339.00,159.92)(12.000,-3.000){2}{\rule{2.891pt}{1.000pt}}
\put(1363,154.92){\rule{5.782pt}{1.000pt}}
\multiput(1363.00,156.92)(12.000,-4.000){2}{\rule{2.891pt}{1.000pt}}
\put(1387,151.42){\rule{6.023pt}{1.000pt}}
\multiput(1387.00,152.92)(12.500,-3.000){2}{\rule{3.011pt}{1.000pt}}
\put(1412,148.42){\rule{5.782pt}{1.000pt}}
\multiput(1412.00,149.92)(12.000,-3.000){2}{\rule{2.891pt}{1.000pt}}
\end{picture}
    \end{center}
    \caption[]{The ratio $X_H/X$ as a function of $m_H$ (GeV), for
      $\tan \beta = 2$ (it scales as $1/\tan^2\beta$).}
     \label{fig:H}
\end{figure}
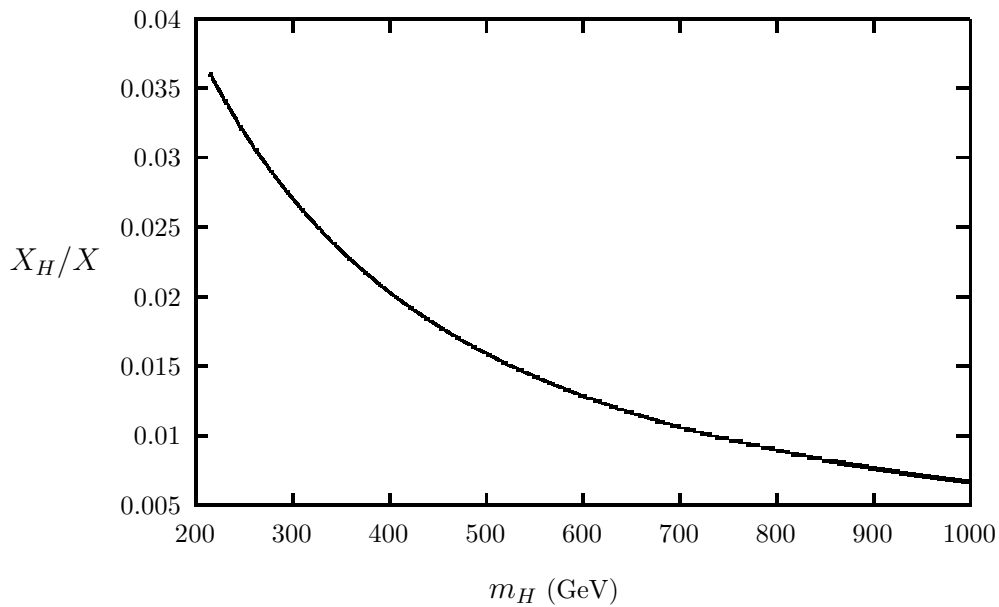

In figures~\ref{fig:0}, \ref{fig:LL}, \ref{fig:LR} and \ref{fig:N} the
ratios  $X_\chi^i/X$ and $X_N/X$
are reported as scatter plots for random values of SUSY parameters in
the ranges given in table~\ref{tab:ranges}. The ratio $X_H/X$ is given
separately in fig.~\ref{fig:H} as a function of $m_H$ for
$\tan\beta=2$ (it scales as $1/\tan^2\beta$). 

\begin{table}
 \begin{center}
 \begin{tabular}{||c|c||}  \hline \hline
 quantity & range\\
 \hline
 $\tan \beta$ & $2\leftrightarrow 6$\\
 $\mu$ &  $(-300\leftrightarrow 300)$ GeV\\
 $M_2$ &  $(100\leftrightarrow 300)$ GeV\\
 $m_{\tilde{t}_R}$ &  $(150\leftrightarrow 300)$ GeV\\
 $m_{\tilde{Q}_L}$ &  $(200\leftrightarrow 500)$ GeV\\
 $m_{\tilde{L}_L}$ &  $(100\leftrightarrow 300)$ GeV\\
 \hline \hline
 \end{tabular}
 \caption[]{Ranges for the SUSY parameters used in the scatter
   plots. Too light charginos are rejected.}
 \label{tab:ranges}
 \end{center}
\end{table} 

Whereas figs.~\ref{fig:0} and~\ref{fig:H} are general in the sense
that their contributions to $r_K e^{-i\theta_K}$ do not depend on the
values of mass insertions, the other ones have to be multiplied by the
corresponding $R$ parameter to get the contribution to $r_K
e^{-i\theta_K}$. On the other hand, an upper limit for the
contribution of the ``constrained'' MSSM, and of any other SUSY
extension of the SM in which all the $R$ parameters are $O(1)$, can be
directly read from these figures. Therefore in this quite large class
of models the neutralino contributions are negligible.  The only
possible non-negligible contributions come from the charged Higgs
exchange, in the ``corner'' of parameter space in which the Higgs mass
is near its lower limit and $\tan\beta$ is very close to one, and from
the term $X^0_\chi$ in chargino exchange. It is interesting to note
that these two contributions have opposite signs and therefore tend to
cancel each other, making the effects very small. This cancellation is
particularly effective if one takes into account the constraints
coming from $b \to s \gamma$, which imply a correlation between the
mass of charginos and charged Higgses. We are therefore able to
conclude that in this class of SUSY models in which the $R$ parameters
are $O(1)$, the effects on $r_K$ are certainly smaller than $10\%$. 
Moreover, if the $R$ parameters are real, as it is the case in the
``constrained'' MSSM, we have $\theta_K=0$.

At this point, we would like to comment on the smallness of the 
chargino and neutralino
contributions, assuming the absence of anomalously
large mass insertions. In the limit of
large squark masses, the box diagrams are suppressed by 
$M^2_Z/m^2_{\tilde{q}}$
relative to penguin diagrams because the $1/M^2_Z$ coming from the $Z$
propagator in the penguin diagrams is replaced by a
$1/m^2_{\tilde{q}}$ coming from the additional superparticle
propagator.  One could then think that the penguin diagrams might
still give contributions not suppressed in that limit. Actually this
is not the case, once again because of gauge invariance, which requires
the presence of at least one gaugino-higgsino flip in the
chargino/neutralino line and an overall $M^2_Z/m^2_{\tilde{q}}$
suppression as in the box-diagram case.

\begin{figure}   
    \begin{center}
       \input{figLL.tex}
    \end{center}
    \caption[]{The ratio $\vert X_\chi^{LL}\vert/X$ as a function of
      $\vert \mu \vert$.} 
     \label{fig:LL}
\end{figure}
\begin{figure}   
    \begin{center}
       \input{figLR.tex}
    \end{center}
    \caption[]{The ratio $\vert X_\chi^{LR}\vert/X$ as a
      function of $\vert \mu \vert$.}
     \label{fig:LR}
\end{figure}
\begin{figure}   
    \begin{center}
       \input{figN.tex}
    \end{center}
    \caption[]{The ratio $\vert X_N\vert/X$ as a function of $\vert
      \mu \vert$.} 
     \label{fig:N}
\end{figure}

\subsubsection{Results in a general SUSY extension of the Standard Model}
\label{sec:genres}

We now turn to the discussion of the effects which can be obtained in
a general SUSY extension of the SM. In this case, we do not assume any
particular model, but instead let the $R$ parameters vary within the
range allowed by the available experimental constraints.

Our numerical analysis uses the constraints on off-diagonal mass
insertions found in ref.~\cite{ggms}. To obtain these
constraints, it has been imposed that the gluino-mediated 
contributions to $K^0-
\bar{K}^0$, $D^0-\bar{D}^0$ and $B^0_d-\bar{B}^0_d$ mixing, as well as to
the decay $b \to s \gamma$, proportional to each single off-diagonal
mass insertion, do not exceed the experimental value. Barring
accidental cancellations, this analysis gives the largest possible
value for the $R$ parameters and consequently for the quantities
as $\theta_K$ or $r_K$ considered in this paper. On the other hand, in
order to simplify our analysis, we have kept the CKM parameters at
their central values (see table \ref{tab:CKM}). In principle, also
these parameters should be varied in the allowed ranges and this
variation should be correlated with the variation of the $R$
parameters in such a manner that the experimental data on all the
quantities listed above are reproduced. Such a very involved numerical
analysis is beyond the scope of this paper. Consequently, the
numerical results presented for this general SUSY extension of the
Standard Model should be considered only as order-of-magnitude estimates.
  
\begin{table}
 \begin{center}
 \begin{tabular}{||c|c||}  \hline \hline
 quantity & value\\
 \hline
 $V_{td}$ & $0.0076 - 0.0029\,i$\\
 $V_{ts}$ & $-0.039 - 0.0006\,i$\\
 \hline \hline
 \end{tabular}
 \caption[]{Values of the relevant CKM matrix elements used in the
   numerical analysis.}
 \label{tab:CKM}
 \end{center}
\end{table} 

We now discuss in more detail the constraints on the off-diagonal mass
insertions which are
relevant in our case. From eqs.~(\ref{CSUSY}) and (\ref{normalization}),
we see that the mass insertions involved in our expressions are
$(m^2_D)_{s_L d_L}$, $(m^2_U)_{s_L d_L}$, $(m^2_U)_{s_L t_R}$ and 
$(m^2_U)_{t_R d_L}$.

Let us start with $(m^2_D)_{s_L d_L}$. From ref.~\cite{ggms} we
immediately get, for degenerate squarks and gluinos,
\bea
{\rm Re} \frac{(m^2_D)_{s_L d_L}}{m^2_{\tilde{d}_L}} <  0.04\,  
\frac{m_{\tilde{d}_L}}{500\, {\rm GeV}}\,,\nn \\
{\rm Im} \frac{(m^2_D)_{s_L d_L}}{m^2_{\tilde{d}_L}} <   0.003 \,
\frac{m_{\tilde{d}_L}}{500\,{\rm GeV}},
\label{ddll}
\eea
which implies the upper limit on $R^D_{s_L d_L}$ reported in
table~\ref{table}.

We now turn to the constraint on $(m^2_U)_{s_L d_L}$.
First of all, following ref.~\cite{pokorski}, 
we note that SU(2) invariance of the soft-breaking
scalar mass matrices implies the following relation:
\be
\label{su2inv}
{\bf m}^{2}_{\tilde{d}_{L}}=V^{\dagger}{\bf m}^{2}_{\tilde{u}_{L}}V.
\ee
On the other hand, we see from eqs.~(\ref{massesrom}) and 
(\ref{massesrom2}) that
\bea
(m^2_U)_{s_L d_L} &=& \left[V^\dagger ({\bf m}^2_U)_{LL} V\right]_{21}=
\left[V^\dagger \left({\bf m}^{2}_{\tilde{u}_{L}} + 
\left({\bf m}^{u}_{D}\right)^2 +
\frac{M_{Z}^{2}}{6}\left(3-4\sin^{2} \Theta_W\right)\cos 2\beta\right) V
\right]_{21}\nn \\
&=&\left[{\bf m}^{2}_{\tilde{d}_{L}}+V^\dagger \left({\bf
      m}^{u}_{D}\right)^2 V \right]_{21}\,,
\label{dudd}
\eea
where we have used eqs.~(\ref{sqmass2}) and (\ref{su2inv}). 
Therefore, we see that, using
eq.~(\ref{ddll}), we can set the following constraint:
\bea
{\rm Re} \frac{(m^2_U)_{s_L d_L}}{m^2_{\tilde{u}_L}} <  0.04\,  
\frac{m_{\tilde{u}_L}}{500\, {\rm GeV}} + 
\frac{{\rm Re} V_{ts}^{*}m^2_t
  V_{td}}{m^2_{\tilde{u}_L}}\,,\nonumber \\
{\rm Im} \frac{(m^2_U)_{s_L d_L}}{m^2_{\tilde{u}_L}} <   0.003 \,
\frac{m_{\tilde{u}_L}}{500\,{\rm GeV}} + 
\frac{{\rm Im} V_{ts}^*m^2_t
  V_{td}}{m^2_{\tilde{u}_L}}\,.
\label{dull}
\eea
This
point was overlooked in ref.~\cite{nir}, in which the authors did
not use eq.~(\ref{su2inv}) to set a constraint on $(m^2_U)_{s_L
d_L}$. Instead, they expanded the product $V^\dagger ({\bf
m}^2_U)_{LL} V$ and used the available constraints on $({\bf
m}^2_U)_{LL}$. This yields a weaker constraint on $(m^2_U)_{s_L
d_L}$, roughly a factor of four larger than the one we
obtain. We will return to this point below.

Concerning the LR and RL mass insertions, things are much more
involved. In this case, we cannot use the SU(2) symmetry to relate the
mass terms in the up sector to the ones in the down sector, and the
limits in the up sector obtained studying gluino-mediated processes
are available only for $(m^2_U)_{c_A u_B}$, with $A,\,B=L,\,R$. On the
other hand, we need to know $(m^2_U)_{t_R d_L}$ and $(m^2_U)_{s_L
  t_R}$. These mass matrix elements could be constrained by studying
chargino contributions to $K^0-\bar{K}^0$ mixing, $B^0_d-\bar B^0_d$
mixing and $b \to s \gamma$ in the mass-insertion approximation.
Unfortunately, there is no such analysis available in the literature,
except for $b \to s \gamma$ \cite{pokorski}, 
and it would go beyond the scope of this paper to discuss these
constraints in detail.  However, an order-of-magnitude estimate of the
limits on these matrix elements can be easily obtained in the
following way.

Let us consider $K^0-\bar K^0$ mixing. The gluino contribution to the
coefficient of the 
$\bar d_L \gamma^\mu s_L\,\bar d_L \gamma_\mu s_L$  operator can be
written as
\begin{equation}
  \label{kkbarglu}
  \frac{\alpha_s^2}{M^2_{\tilde g}} C_{\tilde g} \times \left[
  \frac{(m^2_D)_{s_L d_L}}{\tilde{m}^2}\right]^2,
\end{equation}
where $C_{\tilde g}$ is a dimensionless function of the average squark and
gluino masses, while the chargino contribution includes the following
terms:
\begin{equation}
  \label{kkbarchi}
  \frac{\alpha_W^2}{M^2_\chi} C_{\tilde \chi} \left[\frac{
  (m^2_U)_{s_L t_R} Y_t V_{td}
   + 
  V_{ts}^* Y_t (m^2_U)_{t_R d_L} }{{\tilde{m}^2}}\right]^2.
\end{equation}
Here $C_{\tilde \chi}$ is again a dimensionless function of the average
squark and chargino masses and of the chargino mixing matrices.
We now assume that $\alpha_s^2/M^2_{\tilde g}\,C_{\tilde g} \sim
\alpha_W^2/M^2_\chi\, C_{\chi}$, which is appropriate to get an
order-of-magnitude estimate of the constraints coming from chargino
exchange. Comparing the chargino and gluino
contributions and using the constraint reported in ref.~\cite{ggms}, we
obtain, barring accidental cancellations and
interference effects,
\begin{equation}
  \frac{(m^2_U)_{s_L t_R} Y_t V_{td}}{\tilde{m}^2} \sim  
  \frac{V_{ts}^* Y_t (m^2_U)_{t_R d_L}}{\tilde{m}^2} \sim
  \frac{(m^2_D)_{s_L d_L}}{\tilde{m}^2} < 0.04 \frac{ \tilde{m}}{500 {\rm
      GeV}}.   
  \label{constraintsKK}
\end{equation}
The above equation shows that, due to the presence of the CKM matrix
elements in the relevant couplings, essentially no constraint can be
derived on the 
$(m^2_U)_{t_R d_L}$ and $(m^2_U)_{t_R s_L}$
entries from $K^0-\bar{K}^0$ mixing. Analogously, from $B_d-\bar B_d$
mixing one gets 
\begin{equation}
  \frac{(m^2_U)_{b_L t_R} Y_t V_{td}}{\tilde{m}^2} 
  \sim \frac{V_{tb}^* Y_t (m^2_U)_{t_R d_L}}{\tilde{m}^2} \sim
  \frac{(m^2_D)_{b_L d_L}}{\tilde{m}^2} < 0.1 \frac{\tilde{m}}{500 {\rm
      GeV}}.   
  \label{constraintsBB}
\end{equation}
Therefore, we can set the following approximate limit: 
\be
\left\vert\frac{(m^2_U)_{t_R
  d_L}}{\tilde{m}^2}\right\vert <0.1 \frac{\tilde{m}}{500\,{\rm GeV}}.
\label{consttrdl}
\ee
Constraints coming
from chargino exchange  in $b \to s
\gamma$ have been studied in  ref.~\cite{pokorski}. They report
the following 
constraint: 
\be
\left\vert\frac{(m^2_U)_{t_R s_L}}{\tilde{m}^2}\right\vert < 3\, \left(
\frac{\tilde{m}}{500\, {\rm GeV}}\right)^2.
\label{consttrsl}
\ee

Constraints on the relevant left-right mass insertions can be also
obtained from requiring the absence in the potential of charge and
colour breaking minima or unbounded from below directions\footnote{We
  thank Y. Nir and M. Worah for pointing out to us these constraints,
  which were overlooked in the first version of this work.}. From
ref.\cite{casas} we get:
\be 
\left\vert(m^2_U)_{t_R s_L}\right\vert\,,\,\left\vert(m^2_U)_{t_R
    d_L}\right\vert < m_t \sqrt{2\tilde{m}^2 + m_2^2}\,,
\label{CCB}
\ee
where $m_2$ is the coefficient of the $\vert H_2 \vert^2$ term in the
superpotential. Taking for example $m_2 \simeq \tilde{m}$, one sees
that the FCNC constraint on $(m^2_U)_{t_R d_L}$ in
eq.~(\ref{consttrdl}) is always stronger than the one in
eq.~(\ref{CCB}) in the range of squark masses we consider. On the
other hand, the constraint on $(m^2_U)_{t_R s_L}$ in eq.~(\ref{CCB})
becomes tighter than the FCNC one in eq.~(\ref{consttrsl}) for squark
masses larger than about 300 GeV.

The maximum values for the neutralino and chargino
contributions consistent with present constraints can be estimated 
multiplying the scatter plots in figs.~\ref{fig:0}-\ref{fig:N} by the
upper limits on the relevant $R$ parameters reported in 
table~\ref{table}.

\begin{figure}   
    \begin{center}
       \input{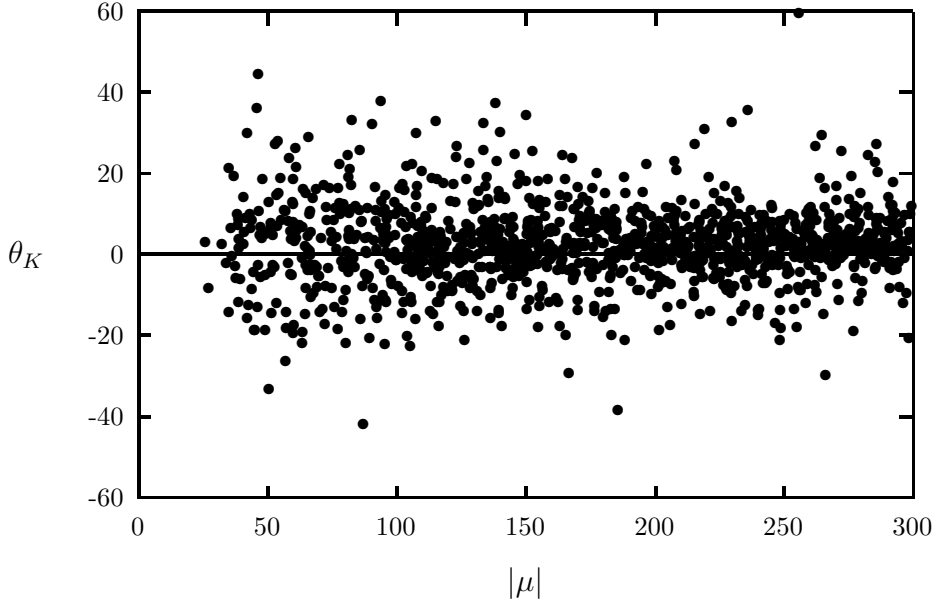}
    \end{center}
    \caption[]{The value of $\theta_K$ (in degrees) obtained by
      varying the $R$ parameters between zero and the upper limits
      given in table \ref{table}, as a function of $\vert
      \mu \vert$ (see the text for details).}
     \label{fig:theta}
\end{figure}

\begin{figure}   
    \begin{center}
       \input{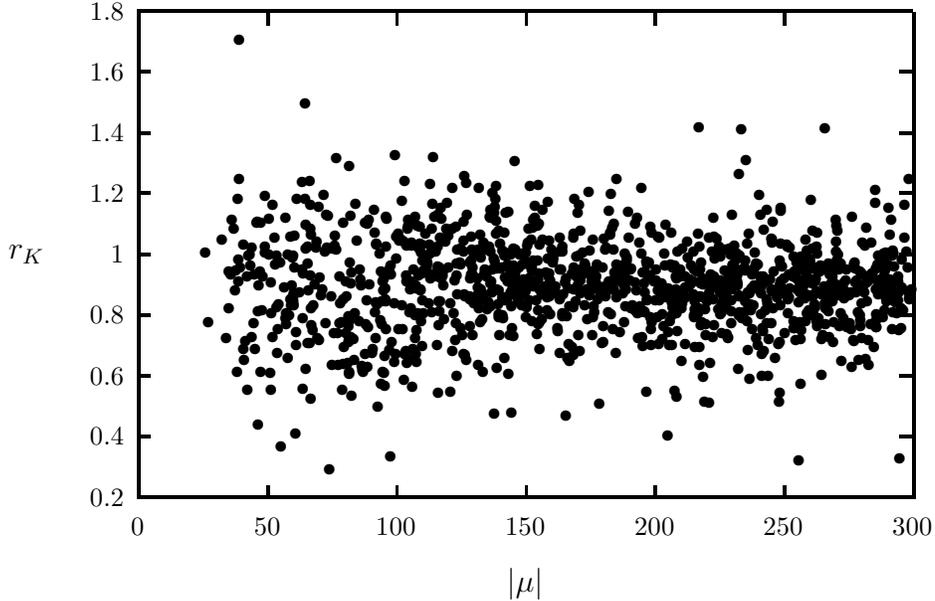}
    \end{center}
    \caption[]{The value of $r_K$ obtained by varying the $R$
      parameters between zero and the upper limits 
      given in table \ref{table}, as a function of $\vert
      \mu \vert$ (see the text for details).}
     \label{fig:rk}
\end{figure}

To provide an order-of-magnitude estimate of the possible size of SUSY
contributions to $r_K$ and $\theta_K$ in this general extension of the
Standard Model, we report in the scatter plots in
figs.~\ref{fig:theta} and \ref{fig:rk} the sum of chargino and
neutralino contributions obtained by varying the R parameters between
zero and the upper limits listed in table \ref{table}. The remaining
SUSY parameters are varied in the ranges reported in table
\ref{tab:ranges}.

We observe that typically 
\be
0.5 < r_K < 1.3 \qquad {\rm and} \qquad -25^0 < \theta_K < +25^0\,,
\label{resgenrt}
\ee
although values outside these ranges, even if less probable, cannot be
excluded. Inspecting eqs.~(\ref{nsin})-(\ref{nbklpn}) we find that
departures from the Standard Model expectations for the branching
ratios in (\ref{brsm}) by factors 2--3 are certainly
possible. Furthermore, in view of substantial values of $\theta_K$,
vanishing or even negative values of $\sin 2(\beta + \theta_K)$ cannot
be excluded. We recall that the standard analysis of the unitarity
triangle gives roughly $\beta \sim 20^0 \pm 8^0$ \cite{BF97}.

Finally we would like to compare our results with those of
ref.~\cite{nir}. In that paper, $K \to \pi \nu \bar \nu$ decays have
been considered in various specific supersymmetric models based on
exact universality, alignment, etc.  It has been found that in these
special models SUSY effects were rather small and the extraction of
$\beta$ from $K \to \pi \nu \bar \nu$ in the presence of SUSY was
possible. On the other hand, the authors of ref.~\cite{nir} stressed
that in generic supersymmetric models large SUSY effects are possible.
We agree with this statement, yet we disagree with the manner this
conclusion has been reached. Nir and Worah attribute these possibly
large effects to flavour violation coming from left-left mass
insertions. As we have demonstrated in (\ref{dull}), a constraint
coming from gluino exchange in $K^0-\bar K^0$ mixing, not considered
in \cite{nir}, excludes such large effects from these insertions. On
the other hand we find that flavour violation in left-right mass
insertions, which was not considered in ref.~\cite{nir}, can give
large contributions to $K \to \pi \nu \bar \nu$, even when available
constraints are taken into account.  Another question is related to
the contribution of box diagrams with chargino exchanges which have
not been considered in ref.~\cite{nir}.  In our opinion the neglect of
these contributions to $X^{LL}_\chi$ cannot be justified.


\subsection{SUSY effects in $B \to \psi K_s$ decays}

We have seen in Sec.~\ref{pheno} that, while in the ``constrained''
MSSM we still expect to be able to measure $\sin 2 \beta$ from $K \to
\pi \nu \bar \nu$ decays with no additional uncertainty with respect
to the Standard Model case, in a general SUSY model large
contributions to $\theta_K$ are possible and therefore the extraction
of the true angle $\beta$ from $K \to \pi \nu \bar \nu$ becomes 
impossible.  Now we would like to briefly comment on SUSY effects in
the measurement of $\sin 2 \beta$ from CP asymmetries in $B \to \psi
K_s$ decays. To this end we use the parameterization (\ref{rdtd}).

Let us consider first the case of the ``constrained'' MSSM. In this
model, no new phase is present ($\theta_d=0$) and therefore the CP
asymmetry $a_{\psi K}$ still measures $\sin 2 \beta$. Consequently, in
the ``constrained'' MSSM the relations (\ref{kbcon}) and (\ref{gr})
are expected to hold. However, as we already explained in Section
\ref{sec:MI}, the {\em true} value of $\sin 2 \beta$ that one can
obtain from $a_{\psi K_s}$ and $K \to \pi \nu \bar \nu$ will generally
differ from the one obtained using the Standard Model analysis of the
unitarity triangle, which is based on $B^0_{d,s}-\bar{B}^0_{d,s}$
mixings and $\varepsilon_K$ \cite{pokorski,branco}.

If we consider instead a general SUSY model, then large contributions
to $B^0_d - \bar{B}^0_d$ mixing and, as we have seen, to $K \to \pi
\nu \bar \nu$ decays are possible \cite{GNW,ggms}.
Moreover, the flavour-changing mass insertions that enter in these
processes are to a large extent uncorrelated. This means that large
violations of the relations (\ref{kbcon}) and (\ref{gr}) are possible
in this case.

\section{Summary and Conclusions}
\label{sec:concl}

In this paper we have presented a model independent analysis of
new-physics contributions to the very clean rare decays $\kpn$    and
$\klpn$. We have illustrated this analysis by considering a large class
of supersymmetric models. In particular, we have investigated whether
the relation (\ref{kbcon}) between $K \to \pi \nu \bar \nu$ decays and
the CP asymmetry $a_{\psi K_S}$ proposed in \cite{BB96} could be
violated in these models. 

The model independent analysis of $K \to \pi \nu \bar \nu$ decays can
be formulated in terms of two parameters $r_K$ and $\theta_K$ in
analogy to the parameters $r_d$ and $\theta_d$ introduced previously
in the literature in connection with $B^0_d - \bar B^0_d$ mixing.
We have demonstrated in Section \ref{sec:MI} how the parameters $r_K$
and $\theta_K$   can be extracted from future data on Br(\kpnn) and
Br(\klpnn). To this end the model independent analysis of the
unitarity triangle of ref. \cite{GNW} has to be simultaneously
invoked.

Analyzing $K \to \pi \nu \bar \nu$ in a large class of supersymmetric
models by means of the mass-insertion approximation of
ref.~\cite{hall} we arrive at the following conclusions:
\begin{itemize}
\item In the ``constrained'' MSSM, $\theta_d=\theta_K=0$, implying
  that the relation (\ref{kbcon}) is well satisfied in this
  supersymmetric scenario. However, the value of $\sin 2 \beta $
  extracted from $K \to \pi \nu \bar \nu$ and $a_{\psi K_S}$ might
  differ from the one obtained using the Standard Model analysis of
  the unitarity triangle.
\item On the other hand, we have demonstrated explicitly that in a
  more general class of supersymmetric models large deviations from
  the Standard Model values $\theta_K=0$ and $r_K=1$ are possible.
  Typically $0.5 < r_K < 1.3$ and $ -25^0 < \theta_K < +25^0$ implying
  that substantial violations of the relation (\ref{kbcon}) and a
  departure from the Standard Model expectations for Br(\kpnn) and
  Br(\klpnn) by factors 2--3 are certainly possible.  Simultaneously,
  the extraction of the CKM parameters, in particular of the angle
  $\beta$, from $K \to \pi \nu \bar \nu$ alone is no longer possible
  and a more complicated analysis involving other decays is necessary,
  as described in Section \ref{sec:MI}. We point out that, although we
  agree with ref.~\cite{nir} that in general supersymmetric extensions
  of the Standard Model the effects on $r_K$ and $\theta_K$ can be
  large, we disagree on the calculation of these effects. In fact, the
  authors of ref.~\cite{nir} state that flavour violation in the mass
  matrices of left-handed up-type squarks could generate
  contributions to $K \to \pi \nu \bar \nu$ as large as the Standard
  Model ones. However, as we already noticed in Section
  \ref{sec:genres}, there is a constraint coming from gluino exchange
  in $K^0-\bar K^0$ mixing which was not considered in
  ref.~\cite{nir}; taking this constraint into account, the left-left
  contribution turns out to be small, about 10\% of the Standard Model
  one. Furthermore, the contribution to $\theta_K$ is even smaller.
  This can be immediately seen multiplying the results in figure
  \ref{fig:LL} by the upper limit for $R^U_{s_L d_L}$ reported in
  table \ref{table}. On the other hand, we find that flavour violation
  in the left-right mass matrices for up-type squarks, which was
  neglected in ref.~\cite{nir}, can give contributions to $K \to \pi
  \nu \bar \nu$ that are much larger, even when the available
  constraints are taken into account. Thus, large SUSY effects in
  $r_K$ and especially in $\theta_K$ are mostly due to left-right
  flavour-changing mass insertions, not to left-left ones.
\end{itemize}

Independently of what kind of new physics is realized in nature, it
is clear that the theoretically clean $K \to \pi \nu \bar \nu$ decays
in conjunction with the very clean CP asymmetry $a_{\psi K_S}$ have
great potentiality to discover new physics and to show us what this
new physics is. In particular, a substantial violation of the golden
relation (\ref{kbcon}) would not only require a generalization of the
Standard Model but would also tell us that it cannot be the
``constrained'' MSSM.

\section*{Acknowledgments} 
We wish to thank M. Misiak for carefully reading the manuscript. This
work has been supported by the German Bundesministerium f{\"u}r
Bildung and Forschung under contract 06 TM 874 and DFG Project Li
519/2-2. L.S. acknowledges the partial support of Fondazione Angelo
della Riccia, Firenze, Italy. The work of A.R. was partially supported
by the TMR Network ``Physics beyond the Standard Model'' under EEC
contract No. ERBFMRX-CT960090.

\section*{Appendix}

We collect here the expressions for various functions which appeared
in Sect.~\ref{RS}.
The formulas in terms of mass insertions that are shown below are
obtained from the general ones by using the following observation.
Consider a $n\times n$ Hermitian matrix $A = A^0 + A^1$ with $A^0 =
{\rm diag}(a^0_1,\ldots, a^0_n)$. If the unitary matrix $U$
diagonalizes $A$ by $A = U^\dagger {\rm diag}(a_1,\ldots,a_n) U$ and
$f$ is an arbitrary function, at the first order in $A^1$ we have
\begin{equation}
U^\dagger_{ik} f(a_k) U_{kj} = \delta_{ij} f(a^0_i) + A^1_{ij}
F(a^0_i, a^0_j),
\end{equation}
where 
\be 
F(x,y) = \frac{f(x) - f(y)}{x-y}.  
\label{fxy}
\ee 
If the matrix $A$ has  degenerate eigenvalues, the limit for $x\to y$
must be performed in eq.~(\ref{fxy}) and one obtains the derivative of
$f(x)$. 
In our case $A$ is the up or down squark mass matrix, $A^0$ is its
diagonal part and $f$ is the appropriate loop function.  We are
performing an expansion at the lowest order in $A^1$. Note that in
this way no ${\cal O}(1)$ uncertainties associated with the use of a
``mean mass'' for squarks of different generations are
present. Therefore, we can take into account the effects of a light
stop.

We consider the first two generations to be almost degenerate in each
squark sector.  Each term in eqs.~(\ref{hesusy}) and (\ref{CSUSY}) has a
contribution from penguin and box diagrams, except for $X_H$ where
the box diagram contribution can be safely neglected.
\begin{eqnarray}
X_H(x) &=& - \frac{m^2_t}{4 M_W^2\tan^2\beta} 
x\left(-\frac{\log x}{(x-1)^2} + \frac{1}{x-1} \right) \nonumber \\
(X_\chi^0)_{\rm Pen} &=& - \frac{m_t^2}{8 \sin^2\beta M^2_W} \left[
k\left( \frac{M^2_{\chi_n}}{m^2_{\tilde{t}_R}},
\frac{M^2_{\chi_m}}{m^2_{\tilde{t}_R}} \right)
H_{h_u^+n}^T H_{nW^+}^* 
H_{W^+m}^T H_{mh_u^+}^* \right. \nonumber \\
&\,& \left. - 2 j\left( \frac{M^2_{\chi_n}}{m^2_{\tilde{t}_R}},
\frac{M^2_{\chi_m}}{m^2_{\tilde{t}_R}} \right)
H_{h_u^+n}^T \frac{M_{\chi_n}}{m_{\tilde{t}_R}}
K_{nW^-}
K_{W^-m}^\dagger \frac{M_{\chi_m}}{m_{\tilde{t}_R}}
H_{mh_u^+}^* 
\right]  \nonumber \\
(X_\chi^0)_{\rm Box} &=& - \frac{m_t^2}{2 \sin^2\beta M^2_W}
\frac{M^2_W}{m^2_{\tilde{t}_R}} 
j\left( \frac{M^2_{\chi_n}}{m^2_{\tilde{t}_R}},
\frac{M^2_{\chi_m}}{m^2_{\tilde{t}_R}},
\frac{m^2_{\tilde{e}_L}}{m^2_{\tilde{t}_R}}  \right)
H_{h_u^+n}^T \frac{M_{\chi_n}}{m_{\tilde{t}_R}}
K_{nW^-}
K_{W^-m}^\dagger \frac{M_{\chi_m}}{m_{\tilde{t}_R}}
H_{mh_u^+}^* \nonumber \\
(X_\chi^{LL})_{\rm Pen} &=& \frac{1}{4} \left[
k\left( \frac{M^2_{\chi_n}}{m^2_{\tilde{u}_L}},
\frac{M^2_{\chi_m}}{m^2_{\tilde{u}_L}}, 1  \right)
H_{W^+n}^T H_{nh_u^+}^* 
H_{h_u^+m}^T H_{mW^+}^* \right. \nonumber \\
&\,& \left. - 2 j\left( \frac{M^2_{\chi_n}}{m^2_{\tilde{u}_L}},
\frac{M^2_{\chi_m}}{m^2_{\tilde{u}_L}}, 1  \right)
H_{W^+n}^T \frac{M_{\chi_n}}{m_{\tilde{u}_L}}
K_{nh_d^-}
K_{h_d^-m}^\dagger \frac{M_{\chi_m}}{m_{\tilde{u}_L}}
H_{mW^+}^* 
\right] \nonumber \\
(X_\chi^{LL})_{\rm Box} &=& -\frac{M^2_W}{m^2_{\tilde{u}_L}}
j\left( \frac{M^2_{\chi_n}}{m^2_{\tilde{u}_L}},
\frac{M^2_{\chi_m}}{m^2_{\tilde{u}_L}},
\frac{m^2_{\tilde{e}_L}}{m^2_{\tilde{u}_L}}, 1  \right)
H_{W^+n}^T \frac{M_{\chi_n}}{m_{\tilde{u}_L}}
K_{nW^-}
K_{W^-m}^\dagger \frac{M_{\chi_m}}{m_{\tilde{u}_L}}
H_{mW^+}^* \nonumber \\
(X_\chi^{LR})_{\rm Pen} &=& -\frac{m^2_t}{4\sqrt{2} \sin\beta M^2_W}
\frac{M_W}{m_{\tilde{u}_L}} \left[
k\left( \frac{M^2_{\chi_n}}{m^2_{\tilde{u}_L}},
\frac{m^2_{\tilde{t}_R}}{m^2_{\tilde{u}_L}}, 1  \right)
H_{W^+n}^T H_{nh_u^+}^* \right. \nonumber \\
&\,& + 2 j\left( \frac{M^2_{\chi_n}}{m^2_{\tilde{u}_L}},
\frac{M^2_{\chi_m}}{m^2_{\tilde{u}_L}} ,
\frac{m^2_{\tilde{t}_R}}{m^2_{\tilde{u}_L}} \right)
H_{W^+n}^T \frac{M_{\chi_n}}{m_{\tilde{u}_L}}
K_{nW^-}
K_{W^-m}^\dagger \frac{M_{\chi_m}}{m_{\tilde{u}_L}}
H_{mh_u^+}^* \nonumber \\
&\,& \left. - k\left( \frac{M^2_{\chi_n}}{m^2_{\tilde{u}_L}},
\frac{M^2_{\chi_m}}{m^2_{\tilde{u}_L}} ,
\frac{m^2_{\tilde{t}_R}}{m^2_{\tilde{u}_L}} \right)
H_{W^+n}^T H_{nW^+}^* 
H_{W^+m}^T H_{mh_u^+}^*
\right] \nonumber \\
(X_\chi^{LR})_{\rm Box} &=& \frac{m^2_t}{\sqrt{2} \sin\beta M^2_W}
\frac{M_W^3}{m_{\tilde{u}_L}^3} 
j\left( \frac{M^2_{\chi_n}}{m^2_{\tilde{u}_L}},
\frac{M^2_{\chi_m}}{m^2_{\tilde{u}_L}}, 
\frac{m^2_{\tilde{t}_R}}{m^2_{\tilde{u}_L}},
\frac{m^2_{\tilde{e}_L}}{m^2_{\tilde{u}_L}} \right)
H_{W^+n}^T \frac{M_{\chi_n}}{m_{\tilde{u}_L}}
K_{nW^-}
K_{W^-m}^\dagger \frac{M_{\chi_m}}{m_{\tilde{u}_L}}
H_{mh_u^+}^* \nonumber \\
(X_N)_{\rm Pen} &=&
j\left( \frac{M^2_{N_n}}{m^2_{\tilde{d}_L}},
\frac{M^2_{N_m}}{m^2_{\tilde{d}_L}}, 1 \right)
U^T_n(d_L) \frac{M_{N_n}}{m_{\tilde{d}_L}}
\left(U_{nh_u^0}U^\dagger_{h_u^0m} -
U_{nh_d^0}U^\dagger_{h_d^0m}\right)
\frac{M_{N_m}}{m_{\tilde{d}_L}} U_m(d_L)^* \nonumber \\
&\,& -\frac{1}{2}
k\left( \frac{M^2_{N_n}}{m^2_{\tilde{d}_L}},
\frac{M^2_{N_m}}{m^2_{\tilde{d}_L}}, 1 \right)
U^T_n(d_L) \left(U_{nh_u^0}^*U^T_{h_u^0m} -
U_{nh_d^0}^*U^T_{h_d^0m}\right) U_m(d_L)^* \nonumber \\
(X_N)_{\rm Box} &=& 
-2 \frac{M_W^2}{m_{\tilde{d}_L}^2} \left[
k\left( \frac{M^2_{N_n}}{m^2_{\tilde{d}_L}},
\frac{M^2_{N_m}}{m^2_{\tilde{d}_L}}, 
\frac{m^2_{\tilde{\nu}_L}}{m^2_{\tilde{d}_L}}, 1 \right)
U^T_n(d_L) U^*_n(\nu_L) U^T_m(\nu_L) U^*_m(d_L)\right. \nonumber \\
&\,& \left. + 2 j\left( \frac{M^2_{N_n}}{m^2_{\tilde{d}_L}},
\frac{M^2_{N_m}}{m^2_{\tilde{d}_L}}, 
\frac{m^2_{\tilde{\nu}_L}}{m^2_{\tilde{d}_L}}, 1 \right)
U^T_n(d_L) \frac{M_{N_n}}{m_{\tilde{d}_L}} U_n(\nu_L)
U_m(\nu_L)^\dagger \frac{M_{N_m}}{m_{\tilde{d}_L}} U^*_m(d_L)
\right]
\end{eqnarray}

We are using the following expressions for the chargino and neutralino
mass matrices:
\begin{eqnarray}
\label{chamass}
M_\chi &=& 
\left(\begin{array}{cc}
M_2 & \sqrt{2} M_W \sin\beta \\
\sqrt{2} M_W \cos\beta & \mu
\end{array}\right) \,,
\\
M_N &=& 
\left(\begin{array}{cccc}
M_1 & 0 & -M_Z \sin\Theta_W \cos\beta &  M_Z \sin\Theta_W \sin\beta \\
0 & M_2 &  M_Z \cos\Theta_W \cos\beta & -M_Z \cos\Theta_W \sin\beta \\
-M_Z \sin\Theta_W \cos\beta & M_Z \cos\Theta_W \cos\beta & 0 & -\mu \\
 M_Z \sin\Theta_W \sin\beta &-M_Z \cos\Theta_W \sin\beta & -\mu & 0
\end{array}\right).\nn
\end{eqnarray}
In the chargino mass matrix rows 1 and 2 correspond respectively to
the $W^-$ and $h_d^-$ current eigenstates, and columns 1 and 2
correspond respectively to the $W^+$ and $h_u^+$ current
eigenstates. In the neutralino mass matrix rows and columns 1, 2, 3
and 4 correspond respectively to the $B$, $W_3$, $h_d^0$ and $h_u^0$
current eigenstates.

The mass eigenstates are obtained by diagonalizing the mass matrices,
using 
\be 
M_\chi = {K}^T M_\chi^D H\,,
\qquad 
M_N = U^T M_N^D U,
\label{KHU} 
\ee 
with $H$, $K$, $U$ unitary matrices and $M^D_\chi =
{\rm diag}(M_{\chi_n})_{n=1,2}$, $M_{\chi_n}>0$ and $M^D_N =
{\rm diag}(M_{N_n})_{n=1,...,4}$, $M_{N_n}>0$.

In the neutralino sector we used the notation
\be
U_n(x) \equiv t_3(x) U_{nW_3} + \tan\Theta_W y(x) U_{nB},
\ee
where $x$ is a generic particle, $t_3(x)$ its third isospin component,
$y(x)$ is its hypercharge, and $U$ is the neutralino mixing matrix
defined in eq.~(\ref{KHU}).

Finally, the loop functions are:
\bea
j(x) = \frac{x\log x}{x-1} &\qquad& 
j(x,y) = \frac{j(x)-j(y)}{x-y} \nonumber \\
j(x,y,z) = \frac{j(x,z) - j(y,z)}{x-y} &\qquad& 
j(x,y,z,t) = \frac{j(x,z,t) - j(y,z,t)}{x-y} \nonumber \\
k(x) = \frac{x^2\log x}{x-1} &\qquad& 
k(x,y) = \frac{k(x)-k(y)}{x-y} \nonumber \\
k(x,y,z) = \frac{k(x,z) - k(y,z)}{x-y} &\qquad &
k(x,y,z,t) = \frac{k(x,z,t) - k(y,z,t)}{x-y}.
\eea

\vfill\eject


\begin{thebibliography}{99}
\bibitem{BS81}
I.I.Y. Bigi and A.I. Sanda, {\sl Nucl. Phys.} {\bf B193} (1981) 85;
\bibitem{BB96}
{ G. Buchalla} and { A.J. Buras}, 
{\sl Phys. Lett.} {\bf B333} (1994) 221;
{\sl Phys. Rev.} {\bf D54} (1996) 6782. 
\bibitem{BB13}
{G. Buchalla and A.J. Buras,}
{\sl Nucl. Phys.} {\bf B398} (1993) 285;
 {\bf B400} (1993) 225;
 {\bf B412} (1994) 106. 
\bibitem{GN}
Y. Grossman and Y. Nir, {\sl Phys. Lett.} {\bf B398} (1997) 163;
\bibitem{GN1}
Y. Grossman, Y.Nir and R. Rattazzi, hep-ph/9701231, to appear in the 
review volume "Heavy Flavours II'', eds. A.J. Buras and M. Lindner
(World Scientific, Singapore); Y. Nir, hep-ph/9709301.
\bibitem{BF97}
{ A.J. Buras and R. Fleischer,} hep-ph/9704376, to appear in the 
review volume "Heavy Flavours II'', eds. A.J. Buras and M. Lindner
(World Scientific, Singapore);
A.J. Buras, hep-ph/9711217.
\bibitem{bnl787}
S. Adler et al., hep-ex/9708031.
\bibitem{ags}
L. Littenberg and J. Sandweiss, eds., AGS2000, Experiments for the
21st Century, BNL 52512.
\bibitem{GNW}
Y. Grossman, Y. Nir and M.P. Worah,
{\sl Phys. Lett.} {\bf B407} (1997) 307.
\bibitem{nir}
Y. Nir and M.P. Worah, hep-ph/9711215.
\bibitem{Littenberg}
{ L.S. Littenberg}, {\sl Phys. Rev.} {\bf D39} (1989) 3322.
\bibitem{MP}
{ W. Marciano and Z. Parsa}, {\sl Phys. Rev.} {\bf D53} (1996) 1.
\bibitem{BB2l}
G. Buchalla and A.J. Buras, hep-ph/9707243. 
\bibitem{WO}
{ L. Wolfenstein}, {\sl Phys. Rev. Lett.} {\bf 51} (1983) 1945.
\bibitem{BLO}
{ A.J. Buras, M.E. Lautenbacher and G. Ostermaier,}
{\sl Phys. Rev.} {\bf D50} (1994) 3433.
\bibitem{ciuchini}
M. Ciuchini, E. Franco, G. Martinelli and L. Silvestrini, 
{\sl Nucl. Phys.} {\bf B501} (1997) 271.
\bibitem{CPASYM}
{ M. Gronau and D. London,} {\sl Phys. Rev. Lett.}
 {\bf 65} (1990) 3381.
\bibitem{SNYD}
A. Snyder and H.R. Quinn, {\sl Phys. Rev.} {\bf D48} (1993) 2139;
{ A.J. Buras and R. Fleischer,}
{\sl Phys. Lett.} {\bf B360} (1995) 138;
J.P. Silva and L. Wolfenstein, 
{\sl Phys. Rev.} {\bf D49} (1995) 1151; 
{ A.S. Dighe, M. Gronau and J. Rosner}, 
{\sl Phys. Rev.} {\bf D54} (1996) 3309; 
R. Fleischer and T. Mannel, 
{\sl Phys. Lett.} {\bf B397} (1997) 269;
C.S. Kim, D. London and T. Yoshikawa, hep-ph/9708356.
\bibitem{bertolini1}
S. Bertolini and A. Masiero, {\sl Phys. Lett.} {\bf B174} (1986) 343.
\bibitem{giudice}
G. Giudice, {\sl Z. Phys.} {\bf C 34} (1987) 57.
\bibitem{bertolini}
{S. Bertolini, F. Borzumati, A. Masiero and G. Ridolfi,}
{\sl Nucl. Phys.} {\bf B353} (1991) 591.
\bibitem{bigi}
I. Bigi and F. Gabbiani, {\sl Nucl. Phys.} {\bf B367} (1991) 3.
\bibitem{couture}
{G. Couture and H. K{\"o}nig,} {\sl Z. Phys.} {\bf C 69} (1995) 167.
\bibitem{hall}
{L.J. Hall, V.A. Kostelecky and S. Raby,} 
{\sl Nucl. Phys.} {\bf B267} (1986) 415.
\bibitem{ggms}
{F. Gabbiani, E. Gabrielli, A. Masiero and L. Silvestrini,}
{\sl Nucl. Phys.} {\bf B477} (1996) 321. 
\bibitem{pokorski}
{M. Misiak, S. Pokorski and J. Rosiek,}
Preprint IFT 3/97, hep-ph/9703442, to appear in the 
review volume "Heavy Flavours II'', eds. A.J. Buras and M. Lindner
(World Scientific, Singapore). 
\bibitem{casas}
J.A. Casas and S. Dimopoulos,  {\sl Phys. Lett.}
{\bf B387} (1996) 107.
\bibitem{branco}
G.C. Branco, G.C. Cho, Y. Kizukuri and N. Oshimo, {\sl Phys. Lett.}
{\bf B337} (1994) 316; {\sl Nucl. Phys.} {\bf B449} (1995) 483;
G.C. Branco, W. Grimus and L. Lavoura, {\sl Phys. Lett.}
{\bf B380} (1996) 119; 
A. Brignole, F. Feruglio and F. Zwirner, {\sl Z. Phys.} {\bf C 71}
(1996) 679.  

  \end{thebibliography}
\end{document}